\documentclass[letterpaper,onecolumn,10pt,accepted=2021-05-24]{quantumarticle}
\pdfoutput=1
\usepackage{authblk}

\title{Efficient Bayesian phase estimation using mixed priors}

\author{Ewout van den Berg}
\affiliation{IBM Quantum, IBM T.J.~Watson Research Center, Yorktown Heights, NY, USA}
\orcid{0000-0002-0991-3397}
\email{evandenberg@us.ibm.com}

\usepackage{amsmath,amssymb,amsthm}
\usepackage{color}
\usepackage{graphicx}
\usepackage{multirow}
\usepackage{bm} 
\usepackage{eufrak}
\usepackage[hidelinks]{hyperref}

\usepackage[numbers,sort&compress]{natbib}
\usepackage{calc}

\usepackage{algorithm}
\usepackage{algpseudocode}

\newcommand{\sfrac}[2]{{\ensuremath{\textstyle\frac{#1}{#2}}}}
\newcommand{\ket}[1]{\ensuremath{\vert{#1}\rangle}}

\newcommand{\abs}[1]{\ensuremath{\vert{#1}\vert}}

\newcommand{\half}[0]{\sfrac{1}{2}}

\newcommand{\norm}[1]{\ensuremath{\Vert{#1}\Vert}}

\definecolor{lightgray}{rgb}{0.94,0.94,0.94}

\definecolor{darkblue}{rgb}{0,0,0.4}

\renewcommand{\vec}[1]{{\color{darkblue}{#1}}} 

\newtheorem{innercustomgeneric}{\customgenericname}
\providecommand{\customgenericname}{}
\newcommand{\newcustomtheorem}[2]{%
  \newenvironment{#1}[1]
  {%
   \renewcommand\customgenericname{#2}%
   \renewcommand\theinnercustomgeneric{##1}%
   \innercustomgeneric
  }
  {\endinnercustomgeneric}
}

\newcustomtheorem{LabelTheorem}{Theorem} 

\begin{document}

\maketitle

\begin{abstract}
  We describe an efficient implementation of Bayesian quantum phase
  estimation in the presence of noise and multiple eigenstates. The
  main contribution of this work is the dynamic switching between
  different representations of the phase distributions, namely
  truncated Fourier series and normal distributions. The
  Fourier-series representation has the advantage of being exact in
  many cases, but suffers from increasing complexity with each update
  of the prior. This necessitates truncation of the series, which
  eventually causes the distribution to become unstable. We derive
  bounds on the error in representing normal distributions with a
  truncated Fourier series, and use these to decide when to switch to
  the normal-distribution representation. This representation is much
  simpler, and was proposed in conjunction with rejection filtering
  for approximate Bayesian updates. We show that, in many cases, the
  update can be done exactly using analytic expressions, thereby
  greatly reducing the time complexity of the updates. Finally, when
  dealing with a superposition of several eigenstates, we need to
  estimate the relative weights. This can be formulated as a convex
  optimization problem, which we solve using a gradient-projection
  algorithm. By updating the weights at exponentially scaled
  iterations we greatly reduce the computational complexity without
  affecting the overall accuracy.
\end{abstract}

\section{Introduction}

Phase estimation is an important building block in quantum computing,
with applications ranging from ground-state determination in quantum
chemistry, to prime factorization and quantum
sampling~\cite{ASP2005DLHa,SHO1997a,TEM2011OVPa}.  In the ideal
setting we assume that the quantum system can be initialized to an
eigenstate $\ket{\phi}$ of a known unitary $U$, such that
\[
U\ket{\phi} = e^{i\phi}\ket{\phi}.
\]
The goal of quantum phase estimation (QPE) is then to estimate the
phase $\phi$. Some of the existing approaches include quantum Fourier
based phase estimation~\cite{ABR1999La,CLE1998EMMa}, iterative phase
estimation~\cite{KIT1995a,SVO2014HFa}, and other methods including
robust-phase estimation, time-series analysis, and integral
kernels~\cite{SOM2019a,ROG2020a,KIM2015LYa}.  In practice, there are
several factors that complicate the problem. First, it may not be
possible to initialize the state exactly to a single eigenstate. This
could be because the state is perturbed by noise, or simply because
the eigenstate is unknown. The latter case arises, for instance, in
the ground state determination of molecules in quantum chemistry where
the desired eigenstate can only be approximated. Regardless of the
cause, phase estimation may need to deal with an initial state that is
a superposition of eigenstates:
\begin{equation}\label{Eq:Psi}
\ket{\Psi} = \sum_j  \alpha_j\ket{\phi_j},\quad\mathrm{with}\quad
\sum_{j} \abs{\alpha_j}^2 = 1,
\end{equation}
Second, practical phase-estimation algorithms may also need to deal
different sources of noise present in current and near-term quantum
devices. Bayesian phase estimation~\cite{SVO2014HFa} has been shown to
be particularly well suited for dealing with noise~\cite{WIE2016Ga}
and the presence of multiple eigenstates \cite{OBR2019TTa}.

In this paper we describe an efficient implementation of Bayesian
phase estimation. In Section~\ref{Sec:Bayesian} we describe the
Bayesian approach to quantum phase estimation along with an
explanation of techniques used to implement it. In
Section~\ref{Sec:Proposed} we review some of the difficulties
encountered in existing techniques and provide a detailed description
of the proposed algorithm. This is followed by numerical experiments
in Section~\ref{Sec:Experiments}, and conclusions in
Section~\ref{Sec:Discussion}.

\begin{figure}[t]
\centering
\includegraphics[width=0.85\textwidth]{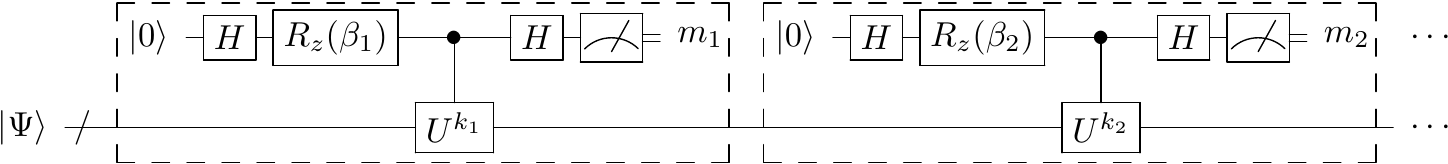}
\caption{The multi-round quantum circuit used in the experiments.}\label{Fig:ExperimentCircuit}
\end{figure}

\section{Bayesian phase estimation}\label{Sec:Bayesian}

We consider Bayesian phase estimation with measurements obtained using
the quantum circuit depicted in Figure~\ref{Fig:ExperimentCircuit}, as
proposed in~\cite{OBR2019TTa}. The circuit consists of a series of
rounds, each parameterized by an integer exponent $k_r$ and phase
shift $\beta_r$. The different rounds in the circuit make up a single
experiment and result in a binary measurement vector $\vec{m}$.
Suppose we are given a quantum circuit of $R$ rounds parameterized by
vectors $\vec{\beta} \in [0,2\pi)^R$ and $\vec{k} \in \mathbb{N}^R$,
then we can denote by
$P_{\vec{k},\vec{\beta}}(\vec{m}\mid \vec{\phi},\vec{w})$ the
probability of observing measurements $\vec{m}$ for given phases
$\vec{\phi}$ and weight vector $\vec{w}$ with entries
$\abs{\alpha_j}^2$. We can define a prior distribution
$P(\vec{\phi},\vec{w})$ over the phases
$\vec{\phi} \in \Phi_{R}=[0,2\pi)^R$, and weights
$\vec{w} \in \Delta$, where $\Delta$ denotes the unit simplex given by
the set
$\{\vec{w} \in \mathbb{R}^R \mid \vec{w} \geq 0, \norm{\vec{w}}_1 =
1\}$.
The posterior distribution then follows from Bayes' theorem, and is
given by
\[
P_{\vec{k},\vec{\beta}}(\vec{\phi}, \vec{w} \mid \vec{m})
= \frac{P_{\vec{k},\vec{\beta}}(\vec{m}\mid \vec{\phi}, \vec{w})\cdot P(\vec{\phi},\vec{w})}
{P_{\vec{k},\vec{\beta}}(\vec{m})},
\]
where
\begin{equation}\label{Eq:P_k_beta_integral}
P_{\vec{k},\vec{\beta}}(\vec{m}) =
\int_{\Phi_R}\int_{\Delta}P_{\vec{k},\vec{\beta}}(\vec{m}\mid
\vec{\phi}, \vec{w})\cdot P(\vec{\phi},\vec{w})\ \!d\vec{w}\ \!d\vec{\phi}.
\end{equation}
Note that throughout the text, the probability density functions are
distinguished by parameter names to keep the notation uncluttered.  By
updating the prior after each set of measurements, we can express the
posterior distribution resulting from the $\ell$-th experiment, with
parameters $\vec{k} = \vec{k}^{(\ell)}$ and
$\vec{\beta}=\vec{\beta}^{(\ell)}$, and measurements
$\vec{m} = \vec{m}^{(\ell)}$, as
\begin{equation}\label{Eq:P_n+1}
P^{(\ell+1)}(\vec{\phi},\vec{w}) = 
\frac{P_{\vec{k},\vec{\beta}}(\vec{m}\mid \vec{\phi}, \vec{w})\cdot P^{(\ell)}(\vec{w},\vec{\phi})}
{P_{\vec{k},\vec{\beta}}^{(\ell)}(\vec{m})},
\end{equation}
which implicitly depends on the parameters and measurements from
previous experiments. In the special case where $\ket{\Psi} = \ket{\varphi}$ is a
single eigenphase, the probability of observing a certain measurement
vector $\vec{m}$ is given by
\begin{equation}\label{Eq:ProbMeasurement}
P_{\vec{k},\vec{\beta}}(\vec{m}\mid \varphi)
= \prod_{r=1}^{R}\cos^2\left(\frac{k_r\varphi}{2} + \frac{\beta_r-m_r\pi}{2}\right)
= \prod_{r=1}^{R} \left(\frac{1 + \cos(k_r\varphi + \beta_r-m_r\pi)}{2}\right).
\end{equation}
In the general case, where $\ket{\Psi} = \sum_j \alpha_j \ket{\phi_j}$ is a superposition of
eigenstates, this changes to
\[
P_{\vec{k},\vec{\beta}}(\vec{m} \mid \vec{\phi}, \vec{w}) = \sum_j w_j P_{\vec{k},\vec{\beta}}(\vec{m}\mid \varphi=\phi_j).
\]
It is reasonable to assume, as done in~\cite{OBR2019TTa}, that the
joint prior $P(\vec{\phi},\vec{w})$ can be written in the form
\[
P^{(0)}(\vec{\phi},\vec{w}) = P^{(0)}(\vec{w})\prod_j
P_j^{(0)}(\phi_j).
\]
Updates to the prior for each of the phases $\phi_j$ can then be
obtained from \eqref{Eq:P_n+1} by integrating out the weights and all
phases except $\phi_j$ (denoted by $\vec{\phi}_{\setminus j}$), which
yields
\begin{equation}\label{Eq:IterationProbPhi}
P_j^{(\ell+1)}(\phi_j) 
= \int_{\Phi_{R-1}}\int_{\Delta} P^{(\ell+1)}(\vec{\phi},\vec{w})\ \!d\vec{w}\ \!d\vec{\phi}_{\setminus j}\\
= 
\frac{P_j^{(\ell)}(\phi_j)}{P_{\vec{k},\vec{\beta}}^{(\ell)}(\vec{m})}
\left(
\sum_{\ell\neq j}C_{\ell}^{(\ell)}W_{\ell}^{(\ell)} +
P_{\vec{k},\vec{\beta}}(\vec{m}\mid \phi_j)W_j^{(\ell)}
\right),
\end{equation}
where
\begin{equation}\label{Eq:CW}
C_j^{(\ell)} = \int_{0}^{2\pi} P_j^{(\ell)}(\phi_j)
P_{\vec{k},\vec{\beta}}(\vec{m}\mid \phi_j),
\quad\mbox{and}\quad
W_j^{(\ell)} = \int_{\Delta} w_jP^{(\ell)}(\vec{w}) d\vec{w}.
\end{equation}
Similarly, integrating over all phases $\vec{\phi}$ gives the updated
weight distribution
\begin{equation}\label{Eq:WeightUpdate}
P^{(\ell+1)}(\vec{w})
= \int_{\Phi_R} P^{(\ell+1)}(\vec{\phi},\vec{w})\ \!d\vec{\phi}
= \frac{P^{(\ell)}(\vec{w})}{P_{\vec{k},\vec{\beta}}(\vec{m})}
\left[\sum_j w_j C_j^{(\ell)}\right].
\end{equation}
Finally, it can be verified that the marginal
probability~\eqref{Eq:P_k_beta_integral} can be expressed in terms of
scalars $C_j^{\ell}$ and $W_j^{\ell}$ from equation~\ref{Eq:CW} as
\begin{equation}\label{Eq:Pkbeta}
P_{\vec{k},\vec{\beta}}^{(\ell)}(\vec{m}) = \sum_{j}C_j^{(\ell)}W_j^{(\ell)}.
\end{equation}
In order to implement Bayesian phase estimation we need a suitable
representation for the probability density functions
$P^{(\ell)}(\phi_j)$ for the phases. We look at these next and discuss
the treatment of weights in more detail in Section~\ref{Sec:Weights}.

\subsection{Normal distribution}

One of the simplest ways to represent the prior $P(\phi)$ is based on
the probability-density function of the univariate normal distribution
$\mathcal{N}(\mu,\sigma^2)$:
\[
f_{\mu,\sigma}(x) = \frac{1}{\sigma\sqrt{2\pi}}e^{-\half\left(\frac{x-\mu}{\sigma}\right)^2}.
\]
As phases that differ by integer multiples of $2\pi$ are equivalent, we
can obtain the desired representation by wrapping the normal
distribution around the interval $[0,2\pi)$ to get
\begin{equation}\label{Eq:P_mu,sigma}
f_{\mu,\sigma}^{\circ}(\phi) = f_{\mu,\sigma}(\phi)
 + \sum_{k=1}^{\infty} \left(f_{\mu,\sigma}(\phi + 2\pi  k)
 + f_{\mu,\sigma}(\phi - 2\pi k)\right).
\end{equation}
After acquiring the measurements from an experiment, we would like to
update the prior distributions to $P_j^{(\ell+1)}(\phi_j)$, as given
in~\eqref{Eq:IterationProbPhi}. However, the resulting candidate
distribution will not be in the form of a wrapped normal distribution,
and we must therefore find the best fit. This amounts to finding the
mean $\mu^{(n+1)}$ and standard deviation $\sigma^{(n+1)}$ of the
candidate distribution and using these to define the updated normal
prior. In order to find these parameters, we have to evaluate the
expected value of $e^{i\phi}$ over the distribution $Q_j =
P_j^{(\ell+1)}$, namely
\begin{equation}\label{Eq:ExpectedEiphi}
\langle e^{i\phi}\rangle_{Q_j} :=
\int_{0}^{2\pi} e^{i\phi}Q_j(\phi)d\phi,
\end{equation}
Given this expectation we can obtain the mean and Holevo variance
using (see for example~\cite{HAY2014Ba}):
\begin{equation}\label{Eq:MeanVar}
\mu_j = \arg(\langle e^{i\phi}\rangle_{Q_j}),
\quad\mbox{and}\quad
\sigma_j^2 = \frac{1}{\abs{\langle e^{i\phi}\rangle_{Q_j}}^2} - 1.
\end{equation}
These values are then used to define the updated prior
$P_j^{(\ell+1)} := f_{\mu_j,\sigma_j}^{\circ}$. Priors based on the
normal distribution were used in the context of (noisy) single-round
experiments with a single eigenstate by~\cite{WIE2016Ga}. In that
work, the integral appearing in \eqref{Eq:ExpectedEiphi} is evaluated
approximately using rejection sampling, which requires a potentially
large number of samples to evaluate accurately. In
Section~\ref{Sec:UpdateNormal} we show how this expectation can be
computed much more efficiently.

\subsection{Fourier representation}

A second way to represent distributions is based on the Fourier series
\[
P(\phi) = c_0 + \sum_{k=1} c_k\cos(k\phi) + s_k\sin(k\phi),
\]
where $c_k$ and $s_k$ are scalar coefficients. The mean and variance
of the resulting distribution over $[0,2\pi)$, are conveniently
expressed in terms of the coefficients as $\mathrm{arg}(c_1 + is_1)$
and $(\pi^2(c_1^2 + s_1^2))^{-1} - 1$, respectively~\cite{OBR2019TTa}.
It is well known that products of sine and cosine terms can be
rewritten as sums of individual sine and cosine terms by, possibly
repeated, application of the product-sum formulas
\begin{align}\label{Eq:ProductToSum}
\begin{split}
\cos(\theta)\cos(\varphi) &= \half\cos(\theta-\varphi) + \half\cos(\theta+\varphi)\\
\sin(\theta)\sin(\varphi)  &= \half\cos(\theta-\varphi) - \half\cos(\theta+\varphi)\\
\sin(\theta)\cos(\varphi) &= \half\sin(\theta+\varphi) + \half\sin(\theta-\varphi).
\end{split}
\end{align}
Together with the identities $\cos(-\phi) = \cos(\phi)$ and
$\sin(-\phi) = -\sin(\phi)$ it follows that sums and products of
Fourier series can also be written as Fourier series. By appropriately
adding or subtracting the product-sum formulas
in~\eqref{Eq:ProductToSum} it follows that
\begin{align}\label{Eq:SplitSineCosine}
\begin{split}
\cos(\theta + \varphi) &= \cos(\theta)\cos(\varphi) - \sin(\theta)\sin(\varphi)\\
\sin(\theta + \varphi) & = \sin(\theta)\cos(\varphi) + \cos(\theta)\sin(\varphi).
\end{split}
\end{align}
With these identities we can write the measurement
probability~\eqref{Eq:ProbMeasurement} in the form of a Fourier series:
\begin{align}\label{Eq:MixedSum}
\begin{split}
P_{\vec{k},\vec{\beta}}(\vec{m}\mid \phi)
& \stackrel{\eqref{Eq:ProductToSum}}{=}
 \sum_{k=0}^R
\left(\alpha_{k}^{\cos}\cos(k\phi + \theta_k^{\cos})
+
\alpha_{k}^{\sin}\sin(k\phi + \theta_k^{\sin})
\right)\\
& \stackrel{\eqref{Eq:SplitSineCosine}}{=}
 \sum_{k=0}^R
\left(\bar{\alpha}_{k}^{\cos}\cos(k\phi)
+
\bar{\alpha}_{k}^{\sin}\sin(k\phi)
\right).
\end{split}
\end{align}
The uniform prior $P(\phi) = 1/2\pi$ can be written as a
Fourier-series with $c_0 = 1/2\pi$ and all other coefficients zero.
Since the update rule in~\eqref{Eq:IterationProbPhi} only multiplies
and adds distributions, the resulting phase distributions $P_j$ can be
written exactly as a Fourier
series~\cite{BER2009HBMa,HAY2014Ba,OBR2019TTa}. More generally, it
follows that whenever the initial prior can be expressed as a Fourier
series, all subsequent phase distributions can be expressed as Fourier
series. In order to evaluate the products of distributions
in~\eqref{Eq:IterationProbPhi} we can use the following convenient
update rules:
\begin{align}\label{Eq:UpdateRules}
\begin{split}
\half\left(1+\cos(k\phi + \gamma)\right)\cos(j\phi)
& = \half \cos(j\phi) + \sfrac{1}{4}\cos(\gamma)[\cos((k+j)\phi) + \cos((k-j)\phi)]\\
& \phantom{=\ \half\cos(j\phi)} -
\sfrac{1}{4}\sin(\gamma)[\sin((k+j)\phi) + \sin((k-j)\phi)]\\[6pt]
\half\left(1+\cos(k\phi + \gamma)\right)\sin(j\phi)
&=\half \sin(j\phi) +
\sfrac{1}{4}\cos(\gamma)[\sin((k+j)\phi) - \sin((k-j)\phi)]\\
& \phantom{=\ \half\sin(j\phi)} +
 \sfrac{1}{4}\sin(\gamma)[\cos((k+j)\phi) - \cos((k-j)\phi)].
\end{split}
\end{align}

\begin{figure}[t]
\centering
\begin{tabular}{cc}
\includegraphics[width=0.34\textwidth]{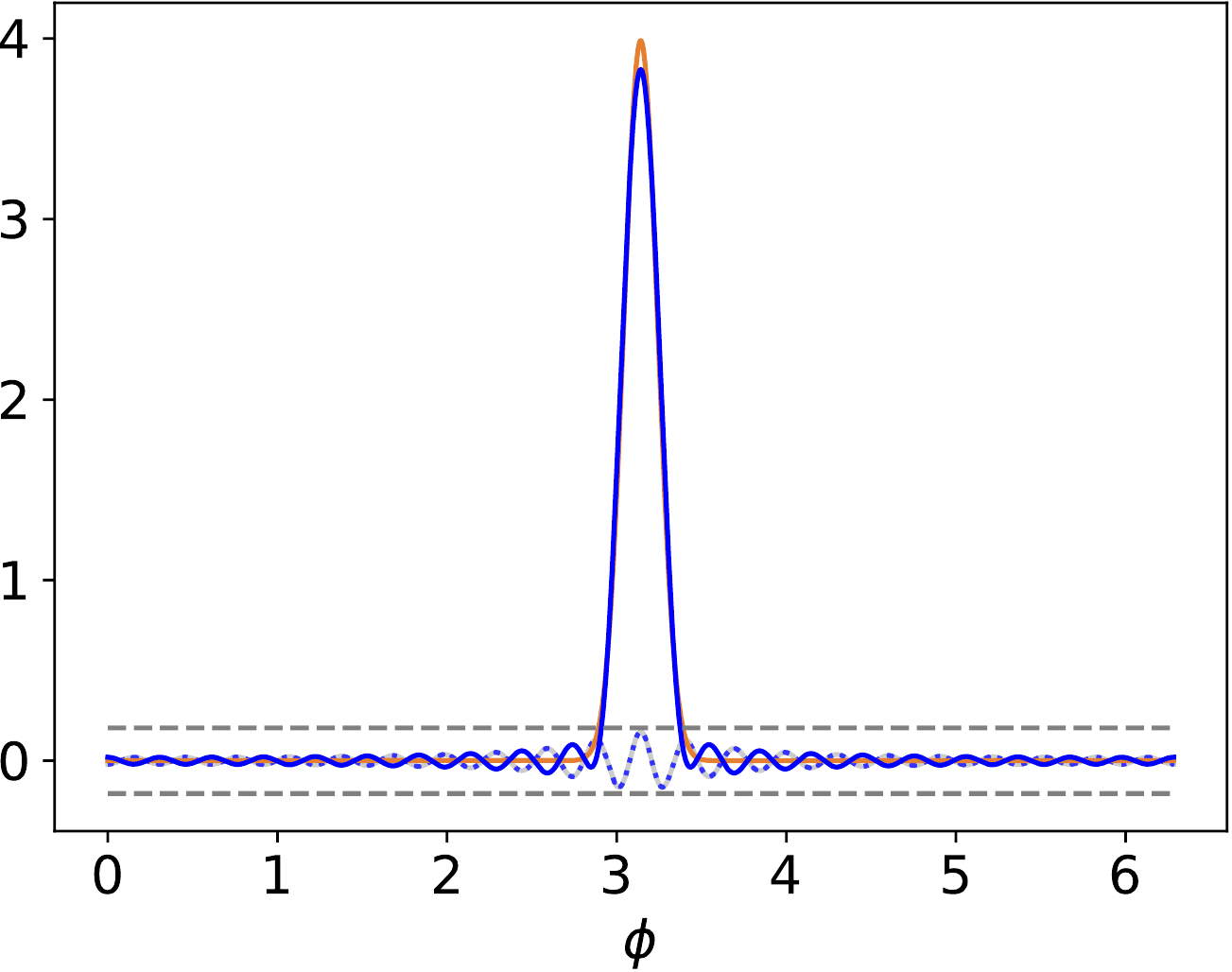}&
\includegraphics[width=0.34\textwidth]{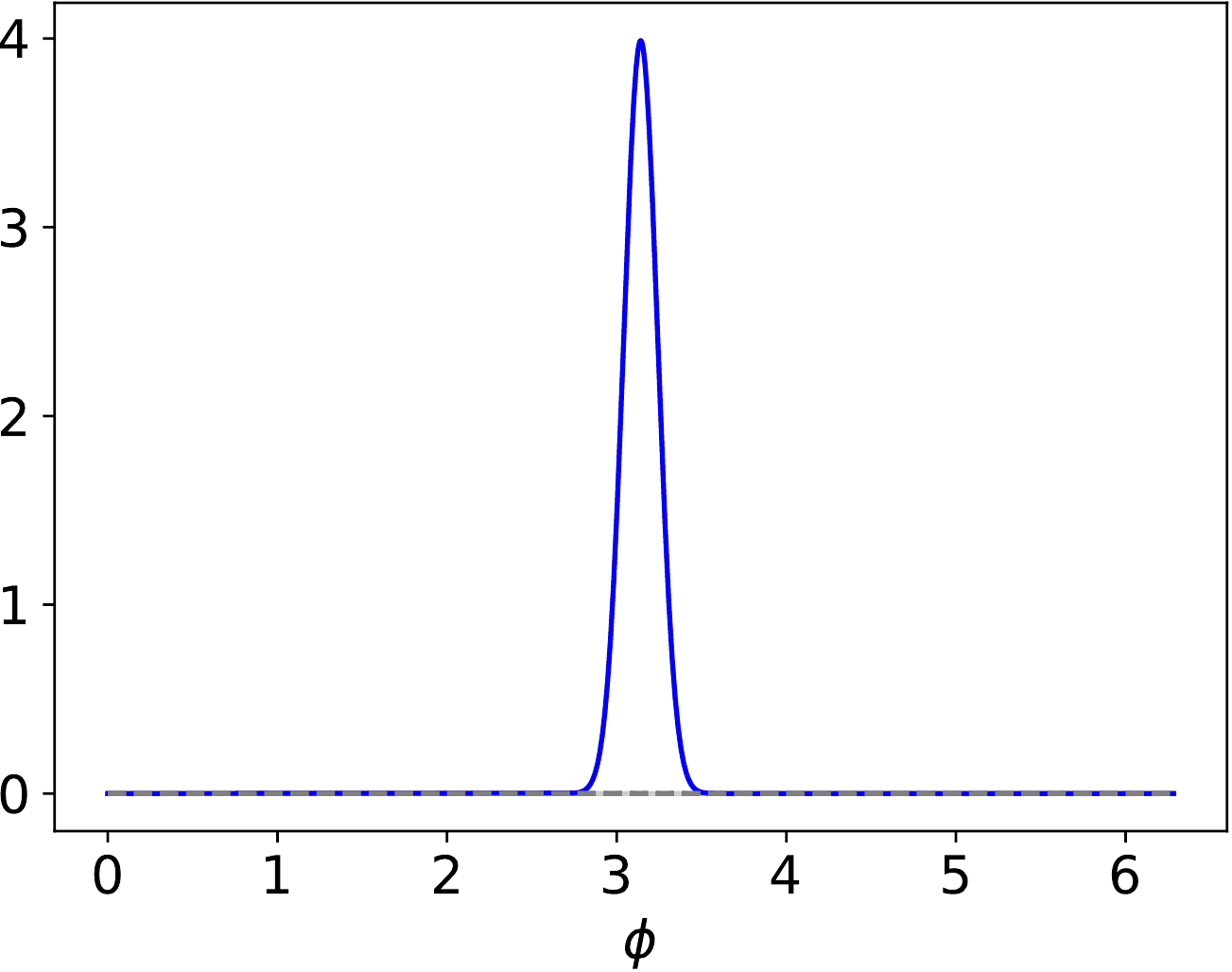}\\
({\bf{a}}) & ({\bf{b}})
\end{tabular}
\caption{Representation of the wrap-around normal distribution with
  $\mu=\pi$ and $\sigma=0.1$ using a truncated Fourier series with
  coefficients up to (a) $n_{\max} = 20$, and (b) $n_{\max} = 50$. The
  dashed and dashed lines respectively indicate the error in the representation
  and our error bound.}\label{Fig:Ringing}
\end{figure}

Given two Fourier series with a finite number of terms $n_1$ and
$n_2$. Then the sum of the two series will have no more than
$\max\{n_1,n_2\}$ terms, whereas the product will have a non-zero term
at the maximum index $n_1+n_2$. When updating the distribution, this
means that each round in the experiment increases the maximum index of
the Fourier-series representation by $k_r$. The number of coefficients
that need to be stored and manipulated therefore keeps growing with
each update, leading to a computational complexity that grows at least
quadratically in the number of rounds. In order to keep the Bayesian
approach practical, we therefore need to limit the number of terms in
the representation~\cite{OBR2019TTa}. However, as the algorithm
progresses, the probability distributions tend to become increasingly
localized and ringing effects, such as those shown in
Figure~\ref{Fig:Ringing}, start to appear as a result of the
truncation. This eventually leads to instability in the distribution,
causing the algorithm to fail. Increasing the number of terms in the
representation can delay but not eliminate this effect. Moreover, from
a computational point of view, we of course would like to keep the
number of terms in the representation as small as possible.

\subsection{Weight updates}\label{Sec:Weights}

Given an initial uniform prior for the weights, we see that
equation~\eqref{Eq:WeightUpdate} evolves to a weighted sum of
Dirichlet distributions, which are of the form:
\[
P_{\vec{\alpha}}(x) = \frac{1}{B(\vec{\alpha})}\prod_i x_i^{\alpha_i-1},
\quad\mbox{with}\quad
B(\vec{\alpha}) = \frac{\prod_i \Gamma(\alpha_i)}{\Gamma(\sum_i \alpha_i)}.
\]
The coefficients $\vec{\alpha}$ for that distribution are initially
zero, but change after each update such that each Dirichlet
distribution has integer parameters $\alpha_i \geq 0$ with
$\norm{\vec{\alpha}}_1 = n$, where $n$ is the update
iteration. Assuming there are $k$ eigenvalues, this gives a total of
${{n+k-1} \choose {k-1}} = \mathcal{O}(n^k)$ terms to represent the
distribution exactly, which becomes impractical for even modest values
of $n$ and $k$. An alternative approach, taken in \cite{OBR2019TTa},
is to choose the weights $\vec{w}$ such that they maximize the
log-likelihood of the measurements.  Assuming a uniform prior, this
leads to a convex optimization problem:
\begin{equation}\label{Sec:WeightOptimization}
\vec{w}^{(n)} := \mathop{\mathrm{argmin}}_{\vec{x}\in\Delta}\
-\frac{1}{n}\sum_{\ell=1}^{n} \log(\langle C^{(\ell)}, \vec{x}\rangle),
\end{equation}
where $C^{(\ell)}$ denotes a vector of the coefficients $C_j^{(\ell)}$
from \eqref{Eq:CW} at iteration $\ell$. The normalization by $1/n$ is
included to keep the problem scale independent of the number of
iterations, but otherwise does not affect the solution.
In~\cite{OBR2019TTa}, this subproblem is solved using sequential
least-squares approach for the first thousand iterations, after which
a local optimization method is applied.

\section{Proposed approach}\label{Sec:Proposed}

In the previous section we have seen that, given initial uniform
priors, the phase distributions have exact Fourier representations up
to the point where the number of coefficients exceeds a chosen
maximum. Truncation of the series eventually leads to instability as
the phase distributions become increasingly localized. In order to
prevent the algorithm from breaking down we propose to monitor the
standard deviation of each phase distribution.  Once the standard
deviation falls below a certain threshold we convert the
Fourier-series representation of the distribution to a normal-based
representation. From then on we only update the parameters of the
normal distribution.

At the beginning of the algorithm we need to choose the priors and
initialize the weights. Starting with identical priors and weights for
each distribution leads to a symmetry in which the updates to the
distributions are identical, which means that we effectively only use
a single distribution. To break this symmetry we can initialize the
priors such that they are all distinct. In
Section~\ref{Sec:FourierOfNormal} we show how a wrap-around normal
distribution can be approximated by a finite Fourier series. This
allows us to initialize the phase distributions with relatively flat
normal-like distributions with different mean values. In
Section~\ref{Sec:FourierError} we bound the pointwise error in
representing a normal distribution by a truncated Fourier series. For
a given maximum number of coefficients in the Fourier representation
we can then find the standard deviation for which the error bound
exceeds a given maximum. This standard deviation is then used as the
threshold that triggers conversion from the Fourier to the normal
representation of the phase distribution.  In
Section~\ref{Sec:UpdateNormal} we show that the parameter updates for
the normal distribution can be computed analytically. This means that
we no longer need the rejection sampling procedure used
in~\cite{WIE2016Ga} to approximate the parameters. Finally, in
Section~\ref{sec:WeightUpdates} we propose an efficient approach for
solving the weight optimization problem~\eqref{Sec:WeightOptimization}.

\subsection{Fourier representation of the normal distribution}\label{Sec:FourierOfNormal}

In order to represent a wrapped normal distribution $f_{\mu,\sigma}^{\circ}$
in Fourier form we need to determine the sine and cosine
coefficients. For a term such as $\sin(k\phi)$, this can be done by
evaluating the inner product:
\begin{equation}\label{Eq:AlphaHat}
\int_{0}^{2\pi} f_{\mu,\sigma}^{\circ}(\phi)\sin(k\phi)d\phi
= \int_{-\infty}^{\infty} f_{\mu,\sigma}(\phi)\sin(k\phi)d\phi
= \langle f_{\mu,\sigma},\sin(k\ \!\cdot)\rangle,
\end{equation}
where the first equality follows from the definition of
$f_{\mu,\sigma}^{\circ}$ in~\eqref{Eq:P_mu,sigma} and periodicity of
the sine function. A similar expression holds for the inner product
with cosine terms $\cos(k\phi)$. We can evaluate these expressions using the
Fourier transform of the normal distribution~\cite{BRY1995a,GRA2008a},
which is given by
\[
\hat{f}(t) = \int_{-\infty}^{\infty}f_{\mu,\sigma}(\phi)e^{it\phi}d\phi
= e^{i\mu t}e^{-(\sigma t)^2/2}
\]
Choosing $t = k$ and taking the real and imaginary parts respectively
we obtain:
\begin{align}
\begin{split}
\langle f_{\mu,\sigma},\sin(k\ \!\cdot)\rangle
& = \sin(\mu k)\cdot e^{-(\sigma k)^2/2}\\
\langle f_{\mu,\sigma},\cos(k\ \!\cdot)\rangle
& = \cos(\mu k)\cdot e^{-(\sigma k)^2/2}.
\end{split}
\label{Eq:InnerProductNormalSineCosine}
\end{align}
For the final coefficients we need to make sure that the basis
functions are orthonormal, which is equivalent to scaling the
coefficients by $1/\pi$ for $k\neq 0$ and by $1/2\pi$ for $k=0$, since
\[
\int_{0}^{2\pi}\cos^2(kx)dx = \begin{cases} 2\pi & k=0\\ \pi & \mathrm{otherwise,}\end{cases}\qquad
\int_{0}^{2\pi}\sin^2(kx)dx = \begin{cases} 0 & k=0\\ \pi & \mathrm{otherwise.}\end{cases}
\]

\subsection{Conversion to normal distribution form}\label{Sec:FourierError}

After each experiment we update the Fourier representation of each of
the eigenphases using the rules in \eqref{Eq:UpdateRules}. Truncation
of the coefficients during the update may introduce intermediate
errors, and is therefore best done after the update by discarding the
excessive coefficients. Over the course of successive updates, the
distributions tend to become increasingly peaked. As a results of the
limited number of coefficients, this causes the distributions to
exhibit the ringing effects seen in Figure~\ref{Fig:Ringing}, which
get more and more severe until the distributions blow up and become
meaningless. We can avoid this by monitoring the standard deviation
for each of the distributions and convert a distribution from Fourier
representation to normal form once the standard distribution becomes
too small. We determine the critical standard deviation by considering
the maximum pointwise error of approximating a normal distribution
with a truncated Fourier series. After normalizing the coefficients
found in \eqref{Eq:InnerProductNormalSineCosine} the pointwise error
at any point $\phi$ for $n_{\max} \geq 0$ is given by summation over
the discarded terms:
\begin{align*}
\mathrm{err}(\phi)
& =\sum_{k=n_{\max}+1}^{\infty} (\sin(\mu k)\sin(k \phi) + \cos(\mu k)\cos(k
\phi))\cdot \frac{e^{-(\sigma k)^2/2}}{\pi}\\
& =\sum_{k=n_{\max}+1}^{\infty} \cos(k(\mu-\phi))\cdot \frac{e^{-(\sigma
  k)^2/2}}{\pi}.
\end{align*}
The maximum absolute error occurs at $\phi=\mu$, where
$\cos(k(\mu-\phi)) = 1$ for all $k$. Defining
$\gamma := e^{-\sigma^2/2} < 1$ we therefore have
\begin{subequations}
\begin{align}
\max_{\phi}\ \abs{\mathrm{err}(\phi)}
& = \frac{1}{\pi} \sum_{k=n_{\max}+1}^{\infty}
  e^{-(\sigma k)^2/2}
= \frac{1}{\pi}\sum_{k=n_{\max}+1}^{\infty} \gamma^{k^2}\label{Eq:FourierErrorBoundA}\\
& \leq \frac{1}{\pi}\int_{n_{\max}}^{\infty} \gamma^{x^2}dx
 = \frac{1}{\pi}\left[\frac{\sqrt{\pi}\cdot\mathrm{erf}(\sqrt{-\ln(\gamma)}x)}{2\sqrt{-\ln{\gamma}}}\right]_{n_{\max}}^{\infty}
= \frac{1}{\pi}\left[\frac{\sqrt{\pi}}{\sqrt{2}\sigma}\mathrm{erf}(x\sigma/\sqrt{2})\right]_{n_{\max}}^{\infty}\notag\\
&= \frac{1}{\sigma\sqrt{2\pi}}\left(1 -
  \mathrm{erf}(n_{\max}\sigma/\sqrt{2})\right)
 =\frac{1}{\sigma\sqrt{2\pi}}\mathrm{erfc}(n_{\max}\sigma/\sqrt{2}).\label{Eq:FourierErrorBoundB}
\end{align}
\end{subequations}
We illustration the error bound in Figure~\ref{Fig:Ringing}. Given a
fixed limit $n_{\max}$ on the number of coefficients and a maximum
permitted error $\varepsilon$, we can define the critical value
$\sigma_{\epsilon}(n_{\max})$ as the minimum $\sigma$ that satisfies
\[
\mathrm{erfc}(n_{\max}\sigma/\sqrt{2}) \leq \varepsilon\sigma\sqrt{2\pi}.
\]
A good approximation can be determined efficiently using bisection
search. Once the standard deviation of the Fourier representation
falls below this critical value, we can no longer guarantee that the
representation is accurate, and therefore switch to a normal
representation of the distribution based on the values
in~\eqref{Eq:MeanVar}. Note that the conversion is done independently
for the different distributions.

\subsection{Updating the normal distribution}\label{Sec:UpdateNormal}

In~\cite{WIE2016Ga}, single-round updates to the normal distribution
are done using rejection sampling. It turns out that the integrals
appearing in the update have closed-form solutions that easily
evaluated. We first consider the computation of the $C_j^{(\ell)}$
coefficients appearing in~\eqref{Eq:CW}:
\[
C_j^{(\ell)} = \int P_j^{(\ell)}(\phi_j)P_{\vec{k},\vec{\beta}}(\vec{m}\mid \phi_j).
\]
It follows from~\eqref{Eq:ProbMeasurement} that we can write
$P_{\vec{k},\vec{\beta}}(\vec{m}\mid \phi_j)$ as a weighted sum of
sine and cosine terms with certain coefficients $c_k$ and $s_k$. Given
a wrap-around normal prior $P_j^{(\ell)} = f_{\mu,\sigma}$, it then
follows that
\[
C_j^{(\ell)} = \int_{-\infty}^{\infty} f_{\mu,\sigma}(\phi)\left[
\sum_{k} c_k\cos(k\phi) + s_k\sin(k\phi)
\right]d\phi.
\]
Using the integrals
in~\eqref{Eq:InnerProductNormalSineCosine}, this simplifies to
\[
C_j^{(\ell)} =
\sum_{k} \left(c_k\cos(\mu k) + s_k\sin(\mu
k)\right)e^{-(\sigma k)^2/2}.
\]
Once the $C_j^{(\ell)}$ values have been computed, we can evaluate the
probability $P_{\vec{k},\vec{\beta}}^{(\ell)}(\vec{m})$
using~\eqref{Eq:Pkbeta}. The posterior distribution used to define the
updated prior is then given by equation~\eqref{Eq:IterationProbPhi}:
\[
\tilde{P}_j^{(\ell+1)}(\phi_j)
=
\frac{P_j^{(\ell)}(\phi_j)}{P_{\vec{k},\vec{\beta}}^{(\ell)}(\vec{m})}
\left(
\sum_{i\neq j}C_{i}^{(\ell)}W_{i}^{(\ell)} +
P_{\vec{k},\vec{\beta}}(\vec{m}\mid \phi_j)W_j^{(\ell)}
\right),
\]
where the tilde indicates that this is an intermediate
distribution. The expression within brackets, along with the
normalization factor $1/P_{\vec{k},\vec{\beta}}^{(\ell)}(\vec{m})$ can be
expressed in Fourier representation with certain coefficients $c_k'$
and $s_k'$. Together with the normal prior distribution, we thus have
\[
\tilde{P}_j^{(\ell+1)}(\phi_j)
= 
f_{\mu,\sigma}(\phi_j) \sum_{k} \left(c_k'\cos(k\phi_j) + s_k'\sin(k\phi_j)\right).
\]
To obtain an updated normal distribution we use~\eqref{Eq:MeanVar}, which requires the
evaluation of $\langle e^{i\phi}\rangle_Q$ with $Q = \tilde{P}_j^{(\ell+1)}$:
\[
\langle e^{i\phi}\rangle_Q = \int_{-\infty}^{\infty} 
f_{\mu,\sigma}(\phi) \sum_{k} \left(c_k'\cos(k\phi) +
  s_k'\sin(k\phi)\right)e^{i\phi}.
\]
Writing $e^{i\phi} = \cos(\phi) + i\sin(\phi)$ and defining
coefficients $c_k^{(\cos)}$, $s_k^{(\cos)}$, $c_k^{(\sin)}$, and
$s_k^{(\sin)}$ such that
\begin{align*}
\sum_{k} c_k^{(\cos)}\cos(k\phi) + s_k^{(\cos)}\sin(k\phi)
& =
\cos(\phi) \sum_{k}\left(c_k'\cos(k\phi) + s_k'\sin(k\phi)\right)
\\
\sum_{k} c_k^{(\sin)}\cos(k\phi) + s_k^{(\sin)}\sin(k\phi)
& =
\sin(\phi) \sum_{k}\left(c_k'\cos(k\phi) + s_k'\sin(k\phi)\right),
\end{align*}
we have
\[
\langle e^{i\phi}\rangle_Q
= \left(\sum_k c_k^{(\cos)}\langle f_{\mu,\sigma},\cos(k\ \!\cdot)\rangle +
s_k^{(\cos)}\langle f_{\mu,\sigma},\sin(k\ \!\cdot)\rangle \right)
+ i\left(\sum_k c_k^{(\sin)}\langle f_{\mu,\sigma},\cos(k\ \!\cdot)\rangle +
s_k^{(\sin)}\langle f_{\mu,\sigma},\sin(k\ \!\cdot)\rangle \right).
\]
This expression is easily evaluated using~\eqref{Eq:InnerProductNormalSineCosine}.

\subsection{Updating the weights}\label{sec:WeightUpdates}

Updates of the weights are done by minimizing the negative
log-likelihood function~\eqref{Sec:WeightOptimization}:
\begin{equation}\label{Sec:WeightOptimization2}
\vec{w}^{(n)} := \mathop{\mathrm{argmin}}_{\vec{x}\in\Delta}\ f_{n}(\vec{x})
\quad\mbox{with}\quad
f_{n}(\vec{x}) := -\frac{1}{n}\sum_{\ell=1}^{n} \log(\langle C^{(\ell)}, \vec{x}\rangle).
\end{equation}
A single evaluation of the objective function and gradient at
iteration $n$ has complexity $\mathcal{O}(mn)$, where $m$ is the total
number of eigenphases considered. If we were to solve this problem for
weight updates in every iteration, we would obtain a complexity that
is quadratic in the number of iterations $n$. Although polynomial,
this nevertheless imposes a practical limit on the number of
iterations the algorithm can take. Adding a single vector $C^{(\ell)}$
to the summation after a large number of iterations does little to
change the objective function or gradient. It therefore makes sense to
solve the weights only when a fixed percentage of new vectors has been
added since the previous solve. In other words, we can use an
exponentially spaced grid of iterations at which to compute the
weights. Assuming a fixed maximum number of iterations to solve each
problem, we then obtain a complexity of $\mathcal{O}(mn)$ for all
weight updates combined, which is linear in the total number of
iterations. An alternative approach, taken in~\cite{OBR2019TTa} is to
solve~\eqref{Sec:WeightOptimization2} using sequential least-squares
programming for the first hundred experiments, and switch to
optimization of a quadratic model of the objective function after
that. Given the constant size of the model, this means that the time
complexity also remains linear in $n$.

For the minimization of~\eqref{Sec:WeightOptimization2} we use a
gradient projection algorithm~\cite{BER1999a}. This iterative method
requires an operator for Euclidean projection onto the unit simplex:
\[
\mathcal{P}(\vec{x}) := \mathop{\mathrm{argmin}}_{\vec{v}\in\Delta}\ \norm{\vec{v}-\vec{x}}_2.
\]
Application of the projection operator can be done using a simple
$\mathcal{O}(m\log m)$ algorithm, which works well for small $m$. More
advanced $\mathcal{O}(m)$ algorithms exist~\cite{CON2016a}, but may
have a large constant factor. We implemented the gradient projection
algorithm with curvilinear line search, in which updates are of the form
\[
\vec{x}^{(i+1)} = \vec{x}^{(i+1)} + \vec{d}^{(i)},\quad\mbox{with}\quad
\vec{d}^{(i)}(\alpha) = \mathcal{P}\left(\vec{x}^{(i)} - \alpha \nabla f_n(\vec{x}^{(i)})\right) - \vec{x}^{(i)}.
\]
The line-search parameter $\alpha$ is chosen to satisfy the Armijo
condition~\cite{NOC2006Wa}
\begin{equation}\label{Eq:Armijo}
f_n(\vec{x}^{(i)} + \vec{d}^{(i)}(\alpha)) \leq f_n(\vec{x}^{(i)}) + \gamma\langle \vec{d}^{(i)}(\alpha),
\nabla f_n(\vec{x}^{(i)})\rangle,
\end{equation}
for some small value of $\gamma$, for instance $\gamma = 0.001$. We
use a backtracking line-search procedure in which the line-search
parameter is initialized to $\alpha_k = 1$, and halved as often as
needed to satisfy~\eqref{Eq:Armijo}. The main stopping criterion is
that $\norm{\vec{d}^{(i)}(1)}_2$ is sufficiently small, as this
indicates that either the norm of the gradient is small, or the
negative gradient lies close to the normal cone of the simplicial
constraint. In addition we impose a maximum number of iterations, as
well as a maximum on the number of halving steps in the line-search
procedure.

\subsection{Noise modeling}\label{Sec:Noise}

One of the advantages of Bayesian phase estimation is that common
types of noise can easily be included in the model. One such type of
noise is depolarizing noise, which results in an ideal noise free
measurement with probability $p$, and a uniformly random
measurement with probability $1-p$. For a single-round experiment, the
probability $p$ can be written as $e^{-k_1/k_{\mathrm{err}}}$, where
$k_{\mathrm{err}}$ is a system-dependent
parameter~\cite{OBR2019TTa}. This can be incorporated in the
conditional measurement distribution by defining
\begin{equation}\label{Eq:MeasProbDecoherence}
\bar{P}_{\vec{k},\vec{\beta}}(m \mid \phi_j)
= p \cdot P_{\vec{k},\vec{\beta}}(m \mid \phi_j) + \frac{1-p}{2},
\end{equation}
and computing $C_j$ in~\eqref{Eq:CW} using the updated measurement
probability. For experiments with $R$ rounds such that each round uses
a different auxiliary qubit and all measurements are taken at the very
end, we can generalize this by replacing $k_1$ by the sum of all $k_i$
values in $p$, and replacing the last term
in~\eqref{Eq:MeasProbDecoherence} by $(1-p)2^{-R}$.

Another example of noise that can be included in the probability model
is measurement noise. Here, each of the measurement outcomes $m_i$ is
flipped with some fixed probability $p$ that can be estimated
experimentally. For multi-round experiments this means that the
current state differs from the one expected based on the measurements
so far. When the number of rounds is limited we can evaluate the
measurement probability as
\begin{equation}\label{Eq:MeasProbMeasurementError}
\bar{P}_{\vec{k},\vec{\beta}}(\vec{m} \mid \phi_j) =
\sum_{\vec{m}'\in \{0,1\}^R} P(\vec{m} \mid \vec{m}')
P_{\vec{k},\vec{\beta}}(\vec{m} \mid \phi_j),
\end{equation}
where $\vec{m}'$ represents the measurements if there were no noise,
and
$P(\vec{m} \mid \vec{m}') =
(1-p)^{R-d(\vec{m},\vec{m}'))}p^{d(\vec{m},\vec{m}')}$
is the probability of observing the noisy measurement $\vec{m}$ with
Hamming distance $d(\vec{m},\vec{m}')$ to
$\vec{m}'$. Both~\eqref{Eq:MeasProbDecoherence}
and~\eqref{Eq:MeasProbMeasurementError} can be written as finite
Fourier series and therefore easily fit into the proposed framework.

\section{Numerical experiments}\label{Sec:Experiments}

In this section we compare the performance of the three different
approaches: the normal approach in which all priors take the form of a
normal distribution; the Fourier approach in which the priors are
represented as a truncated Fourier series; and the proposed mixed
approach, in which each of the priors is initially represented as a
truncated Fourier series and converted to normal distribution form
once the standard deviation falls below a given threshold. For all
experiments we choose phase shift values $\beta$ uniformly at random
on the interval $[0,2\pi)$, and initialize the prior distributions as
equally spaced normal distributions with a standard deviation of 3,
either directly, or approximately using the techniques described in
Section~\ref{Sec:FourierOfNormal}. By choosing different mean values
we break the model symmetry, which would otherwise lead to identical
updates to all distributions. All experiments were run on a 2.6 GHz
Intel Core i7 processor with 16 GB of memory. The results in this
section can be reproduced using the code available online at:
\url{https://github.com/ewoutvandenberg/Quantum-phase-estimation}.

\subsection{Fourier representation}\label{Sec:ExperimentsFourierRepr}
\begin{figure}[t!]
\centering
\begin{tabular}{ccc}
\includegraphics[width=0.305\textwidth]{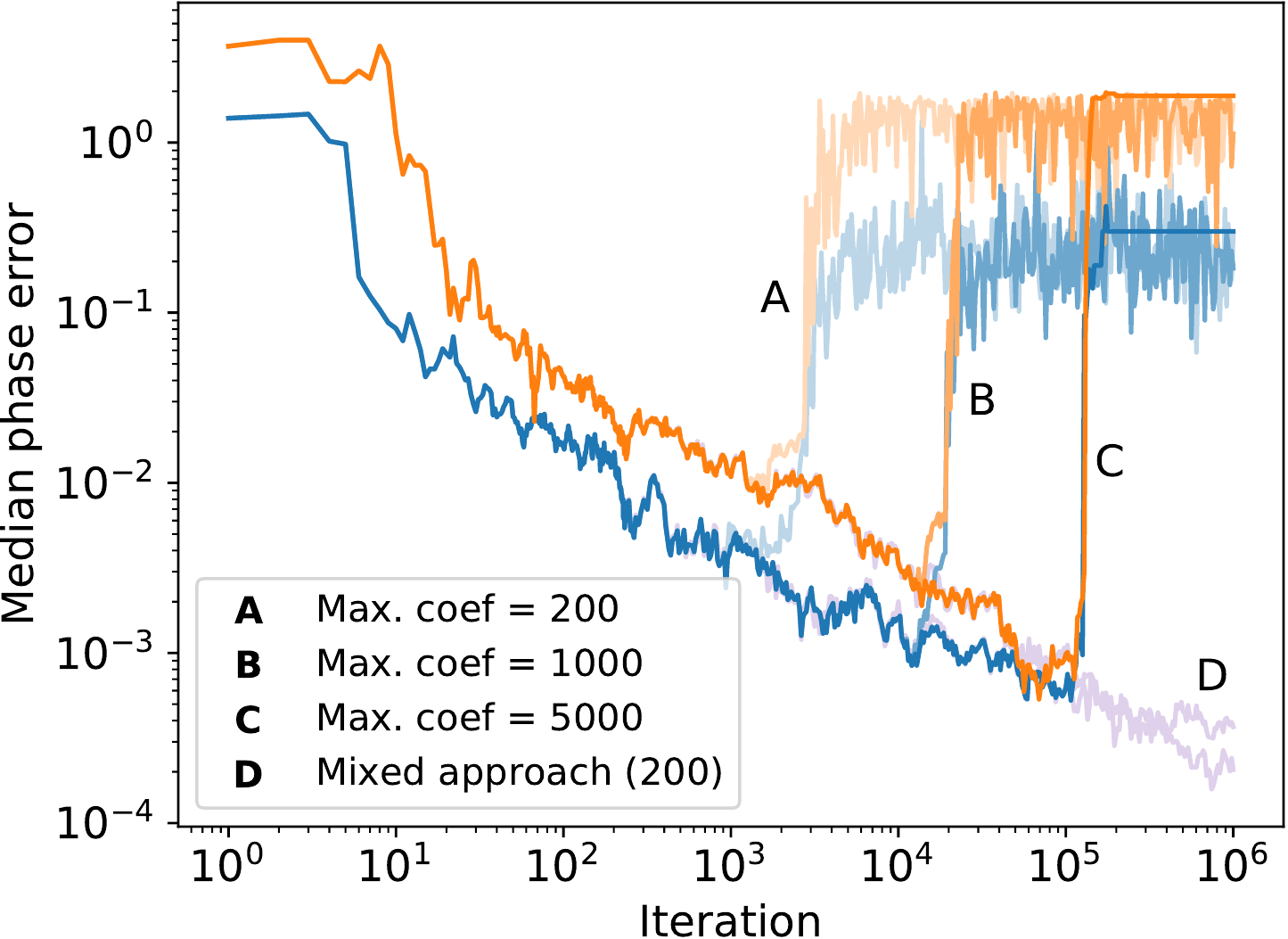}&
\includegraphics[width=0.305\textwidth]{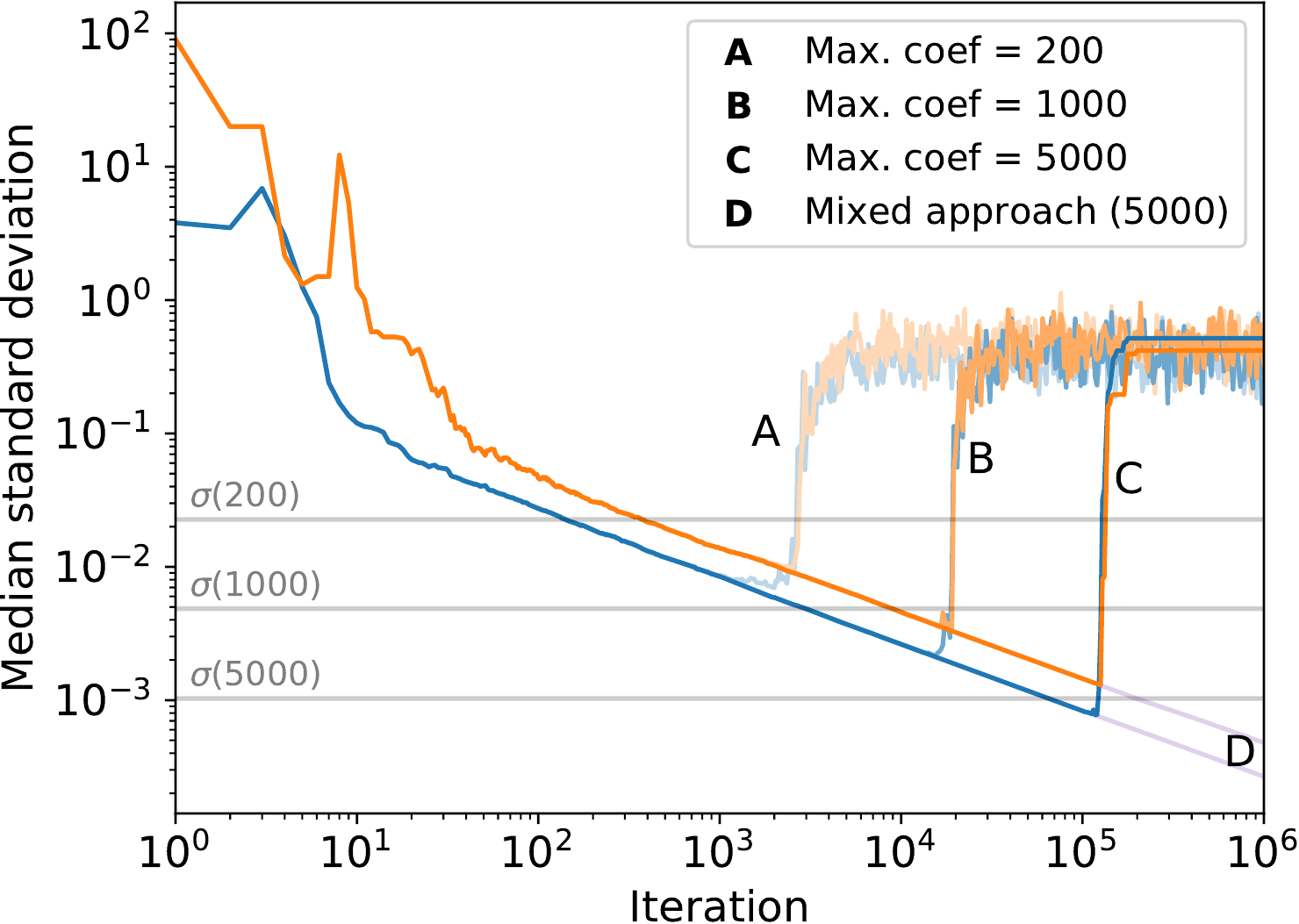}&
\includegraphics[width=0.305\textwidth]{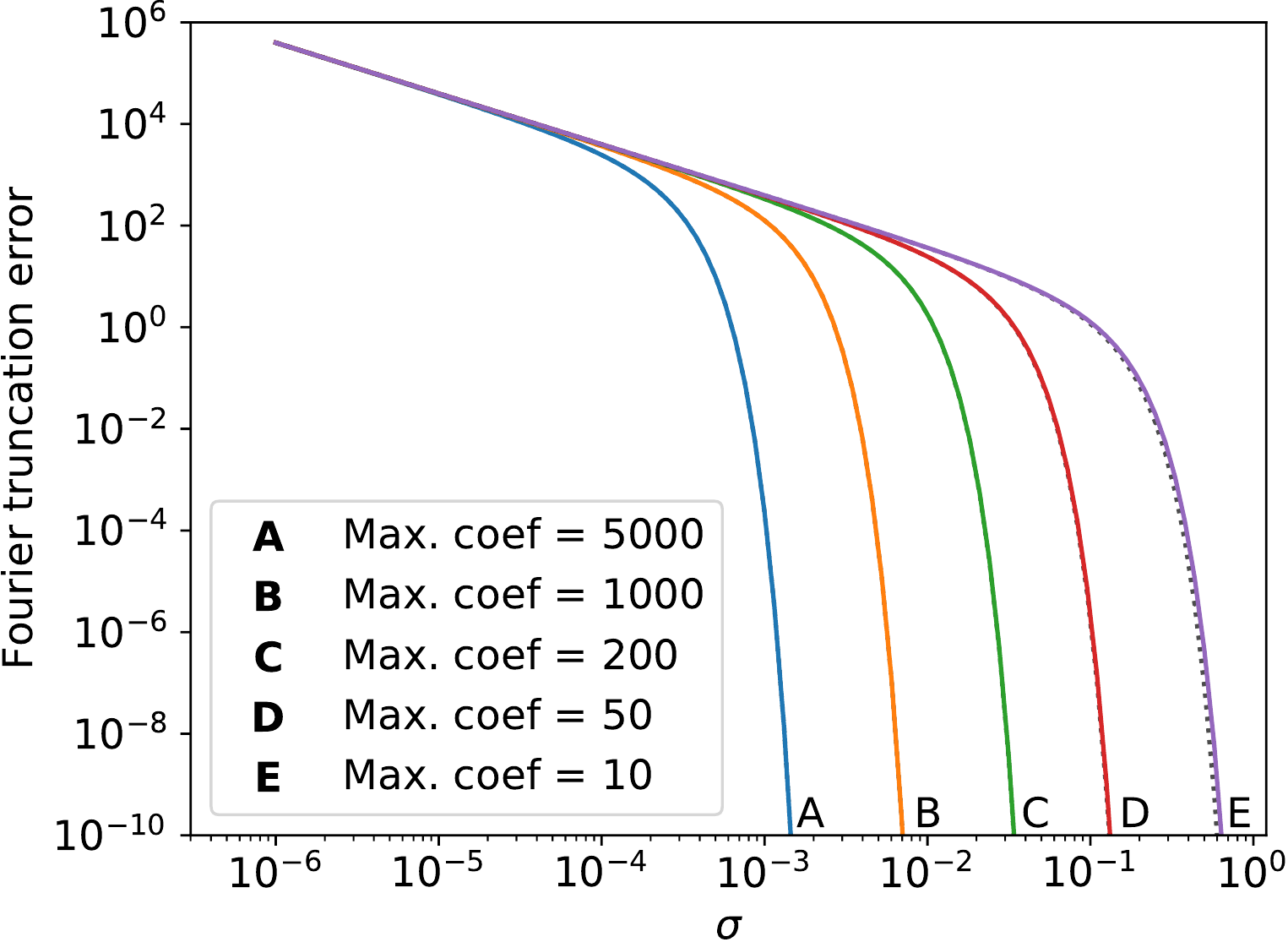}\\
({\bf{a}}) & ({\bf{b}}) & ({\bf{c}}) \\[5pt]
\includegraphics[width=0.305\textwidth]{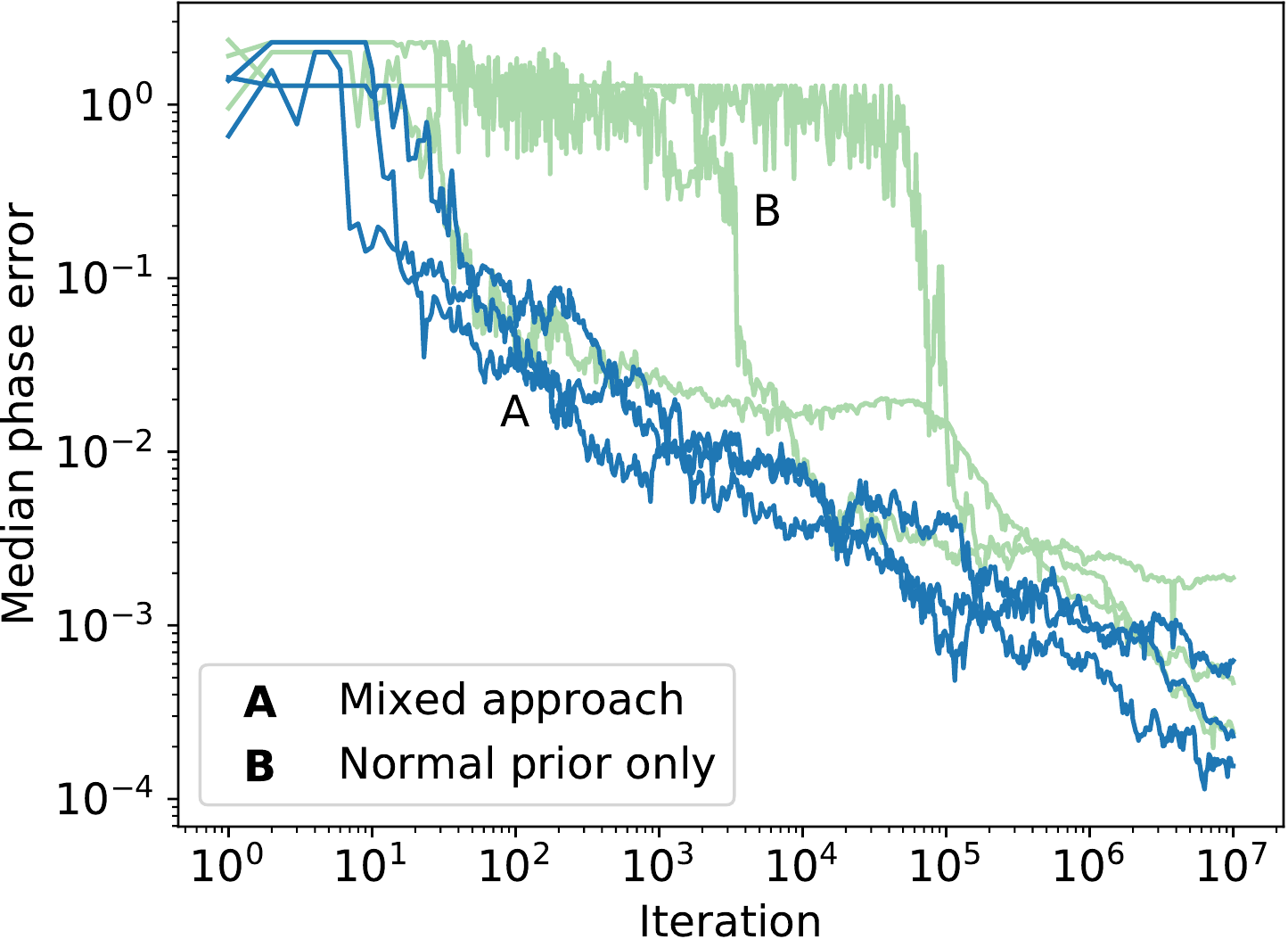}&
\includegraphics[width=0.305\textwidth]{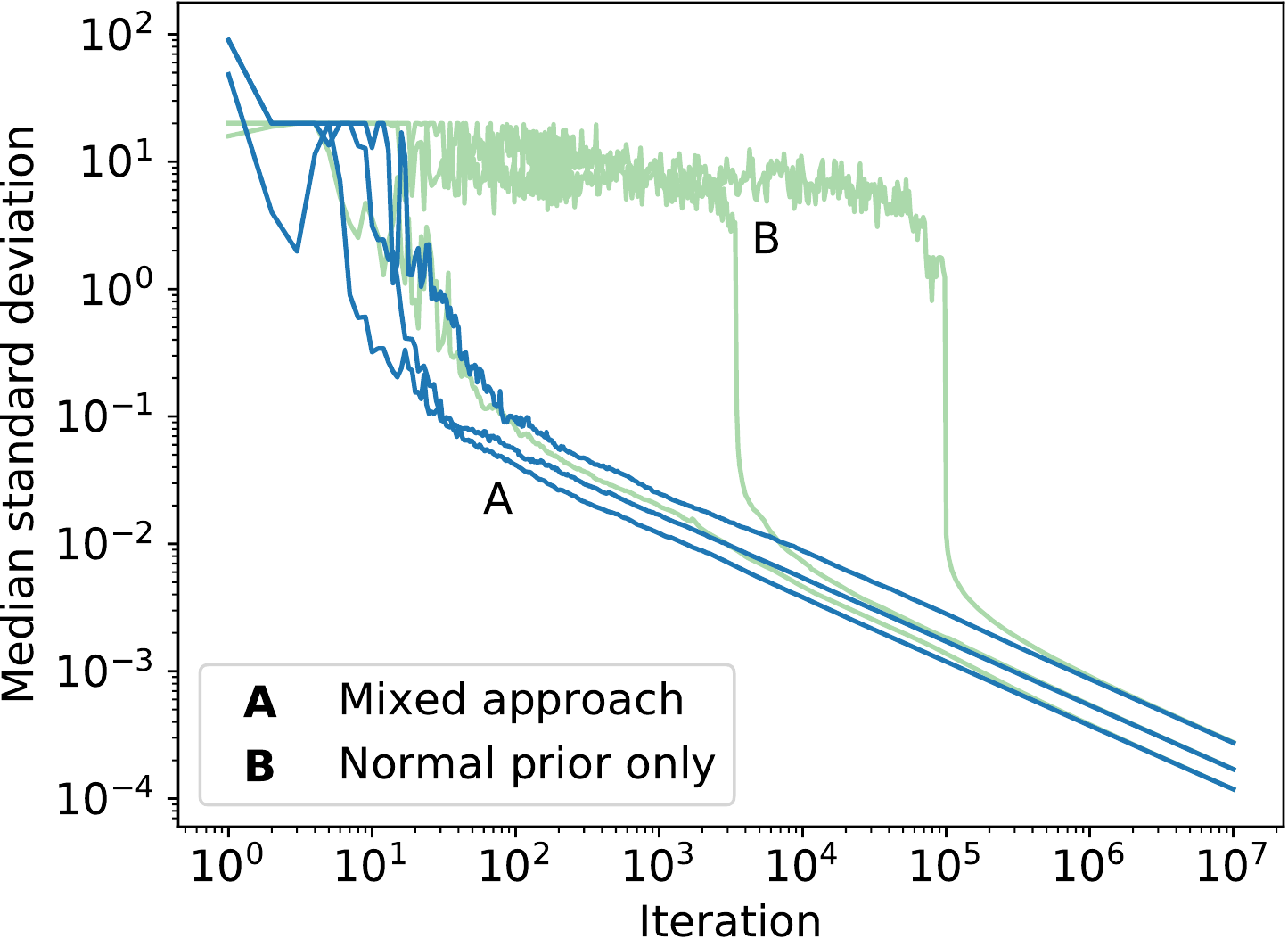}&
\includegraphics[width=0.305\textwidth]{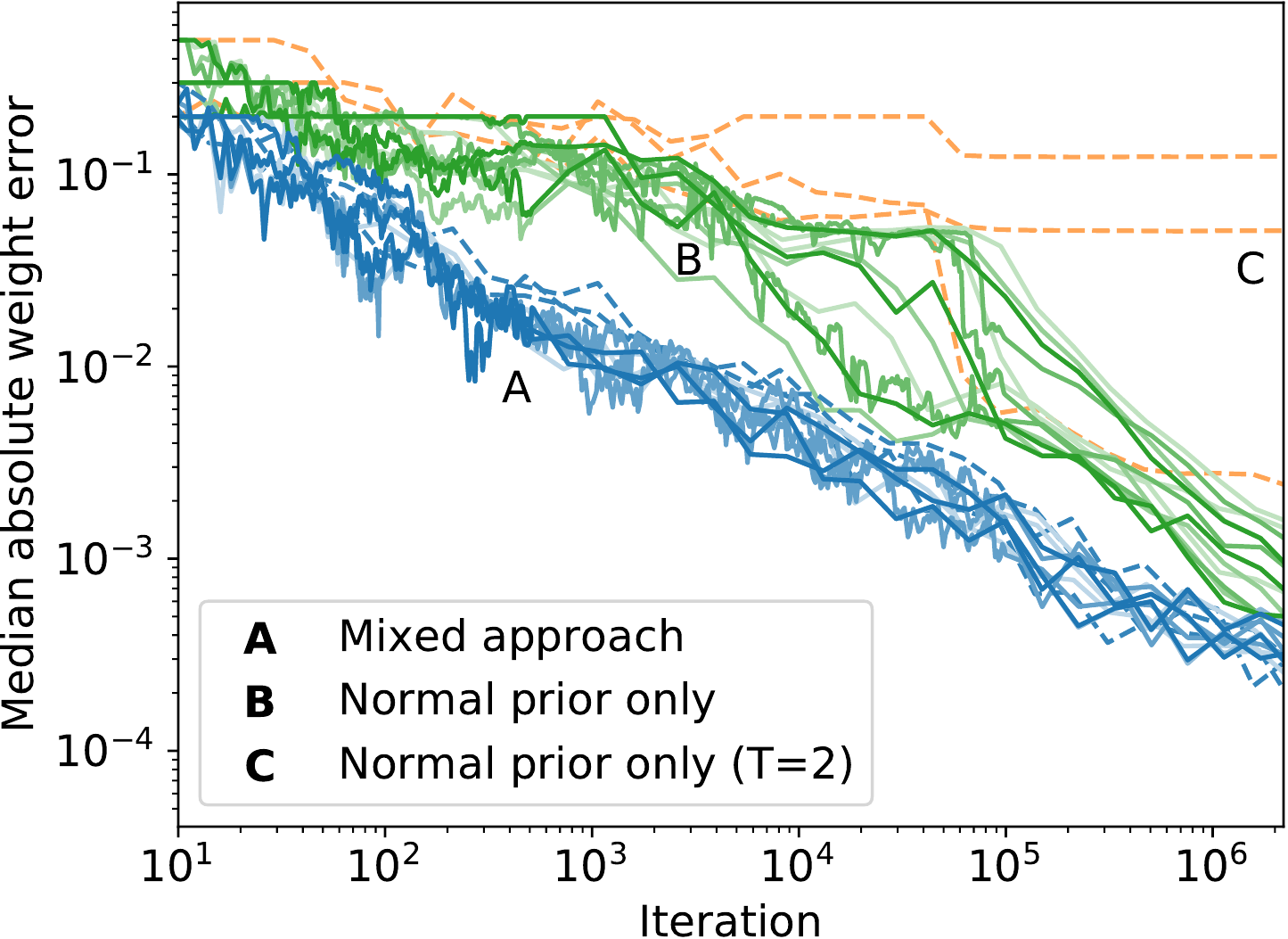}\\
({\bf{d}}) & ({\bf{e}}) & ({\bf{f}}) \\[5pt]
\end{tabular}
\caption{Plots of (a) the median phase error for a two-eigenphase
  problem using Fourier representations with three different values
  for the maximum number of coefficients; (b) the corresponding median
  standard deviation along with the critical $\sigma$ values for
  $\epsilon=10^{-4}$; (c) the Fourier truncation error as a function
  of $\sigma$; (d) the median phase error for three phases using the
  normal-based approach (light green, marker B), and the mixed
  approach (dark blue, marker B), as well as the corresponding (e)
  median standard deviation in the phases, and (f) median absolute
  error in the weights. The curve bundles in plot (f) illustrate the
  use of different schemes to update the weights of the
  distributions. For each approach the different update schemes
  generally perform similarly, and we therefore omit explicit
  labels.}\label{Fig:ExperimentHybrid1}
\end{figure}

As a first experiment we consider the application of the Fourier
approach on a noiseless test problem in which two eigenstates with
eigenphases $\phi_1=2$ and $\phi_2 = 4$ are superimposed with weights
$w_1 = 0.7$ and $w_2 = 0.3$. We model the phase distributions using a
truncated Fourier series and consider truncation after 200, 1000,
and 5000 terms respectively. For the experiments we use a fixed
three-round setup with $\vec{k} = [1,2,5]$, and run 20 independent
trials of $10^6$ iterations each. As the distributions evolve they may
converge to eigenphases corresponding to any of the actual
eigenstates. In order to evaluate the accuracy of the results, we
therefore need to associate each distribution with one of the know
eigenphases. There are several ways in which this matching can be
done. The simple approach we found to work very well, and which we
therefore use throughout, consists of matching the distribution to the
eigenphase that is closest to its mean at a given reference
iteration. We typically perform the matching based on the distribution
at the final iteration, but intermediate distributions are sometimes
used to better illustrate the performance. Applying this procedure to
our current setup with matching done at iteration 100 gives median
phase errors as shown in Figure~\ref{Fig:ExperimentHybrid1}(a). For
each of the three truncation limits we see that the median phase error
initially converges steadily before suddenly diverging rapidly. For
the Fourier representation with 200 terms this divergence happens
after some 2,500 iterations. Using representations with a larger
number of terms helps postpone the point at which divergence sets in,
but does not prevent it.  Matching at different iterations beyond
1,000 confirms that the divergence is inherent in the representation
and not caused by the distributions suddenly converging to different
eigenphases. In Figure~\ref{Fig:ExperimentHybrid1}(b) we plot the
median standard deviation as a function of iteration along with the
critical $\sigma$ values for $\epsilon = 10^{-4}$, as calculated using
the procedure described in Section~\ref{Sec:FourierError}. For
plotting purposes we set the standard deviation to 20 when the weight
of the distribution is zero. For the case where the maximum number of
coefficients is 5,000, we see that the break down occurs very quickly
after the critical standard deviation is reached. To get a better
understanding of the sensitivity of this critical value with respect
to $\epsilon$, we plot the Fourier truncation
error~\eqref{Eq:FourierErrorBoundB} as a function of $\sigma$ in
Figure~\ref{Fig:ExperimentHybrid1}(c). The dashed line in the plot
gives a more accurate bound on the error obtained by explicitly
summing the terms in \eqref{Eq:FourierErrorBoundA} up to 100,000, and
using the complementary error-function (erfc) bound beyond that. The
fact that the dashed line is barely visible shows that the bound in
\eqref{Eq:FourierErrorBoundB} is reasonably tight, certainly for our
purposes. In addition, it can be seen that the truncation error
exhibits a sharp increase over a narrow range of $\sigma$ values. This
means that the critical value is not very sensitive to the choice of
$\epsilon$; choosing $10^{-4}$ or $10^{-6}$ will give very similar
critical values. On the other hand, the sharp increase also does mean
that any further decrease in the standard deviation beyond this point
results in rapidly increasing errors. The critical sigma values can be
seen to decrease with the number of coefficients in the
representation, and it may therefore appear that choosing a large
number of coefficients solves the problem. However, the computational
complexity of updating the representation increases with the number of
terms. For instance, for the three maximum values used in our examples
we recorded average runtimes of 68, 323, and 1552 seconds per
trial. Further increasing the number of terms will become
computationally expensive, and we therefore conclude that a pure
Fourier-based implementation is not practical in general.

\subsection{Normal and mixed representations}\label{Sec:ExperimentsNormalMixed}

We now move on to another experiment in which we compare the normal
and mixed approach. For the mixed approach we use a Fourier
representation with 200 coefficients and a critical standard deviation
corresponding to $\epsilon = 10^{-4}$.  We maintain the non-adaptive
measurement scheme described above, but change to a new problem with
three phases $\vec{\phi} = [2,4,5]$ and weight values
$\vec{w} = [0.5,0.3,0.2]$.  Given that the number of eigenstates in
the initial state $\ket{\Psi}$ is generally not known, we maintain
five distributions for the phases. For the performance evaluation we
match each distribution to one of the three ground-truth values. The
weight values contributing to each distribution are added up, and the
phases are averaged using the relative weights\footnote{In practice we
  do not know which, if any, of the distributions should be
  merged. One possible approach is compute the likelihood of the
  separate versus combined distributions and find the grouping that
  maximizes the likelihood. For this work we did not pursue any such
  approach.}.  The resulting median phase errors for the collated
distributions are plotted in
Figure~\ref{Fig:ExperimentHybrid1}(d). The normal-based approach has a
slower start than the mixed approach, but eventually catches up after
around $10^5$ iterations. The standard deviation of the distributions
and the convergence of the weights in terms of the median absolute
distance are plotted in Figures~\ref{Fig:ExperimentHybrid1}(e--f). A
closer look at the data indicates that the mixed approach successfully
converged to the correct weights in all twenty trials. The
normal-based approach does so in the majority of trials, but has some
notable deviations, which lead to a deteriorated mean absolute
error. The average runtime of the mixed approach was 70.43 seconds,
which is larger than the average of 15.65 seconds for the normal-based
approach. The difference in runtime is due to the updates of the
Fourier representation in the initial iterations, which are more
expensive. In most of the trials of the mixed approach we found
exactly three distributions with non-negligible weights. For these
distribution, the transition from Fourier representation to normal
distribution occurred on average at iterations 282, 533, and 1111, in
order of decreasing weight of the distribution.

\subsection{Weight updates}

In Section~\ref{sec:WeightUpdates} we proposed a way of reducing the
number of weight updates to ensure that the computational complexity
scales linearly instead of quadratically in the number of
iterations. To validate the approach we evaluate the performance of
the mixed approach with five different weight-update settings. For each
of these we update the weights at every iteration for the first $T$
iterations. After that, the weights are updated only at iterations
that are power-of-two multiples of $T$, namely, $2T$, $4T$, $8T$, and
so on. In the first setting we update the weight at every iteration
for the first $T=10^5$ iterations. This setting therefore closely
resembles the naive approach with a complexity that scales
quadratically in the number of iterations. In the next three settings
we respectively choose $T=2$, $T=64$, and $T=512$. In the fifth
setting we again choose $T=512$, but update the line-search as
described in Section~\ref{sec:WeightUpdates}. Rather than projecting
$x+\alpha d$ for each line-search parameter $\alpha$, we project once
with $\alpha=1$, and perform backtracking line search using only that
point. We evaluate these settings on the problem described in
Section~\ref{Sec:ExperimentsNormalMixed} and plot the resulting mean
absolute weight errors in Figure~\ref{Fig:ExperimentHybrid1}(f). The
different shades of the colors used for the normal and mixed
approaches correspond to the different update schemes. The only
setting that failed to converge was the normal approach combined with
update parameter $T=2$. Aside from this all update schemes performed
comparably for each of the approaches. The only real difference, not
visible in the plot, lies in the runtime: in the mixed approach, the
first setting, in which updates the weights at every iteration for the
first $10^5$ iterations, takes 1040 seconds to complete. Settings two
through four take 472, 587, and 610 seconds, and the final setting
takes 604 seconds. For the normal approach these runtimes are 366,
117, 122, 121, and 126, respectively. The experiments in
Section~\ref{Sec:ExperimentsNormalMixed} were done using $T=512$, and
we will continue to use this value for the remainder of the paper.

\subsection{Design of single-round experiments}

\begin{figure}[t!]
\centering
\begin{tabular}{ccc}
\includegraphics[width=0.305\textwidth]{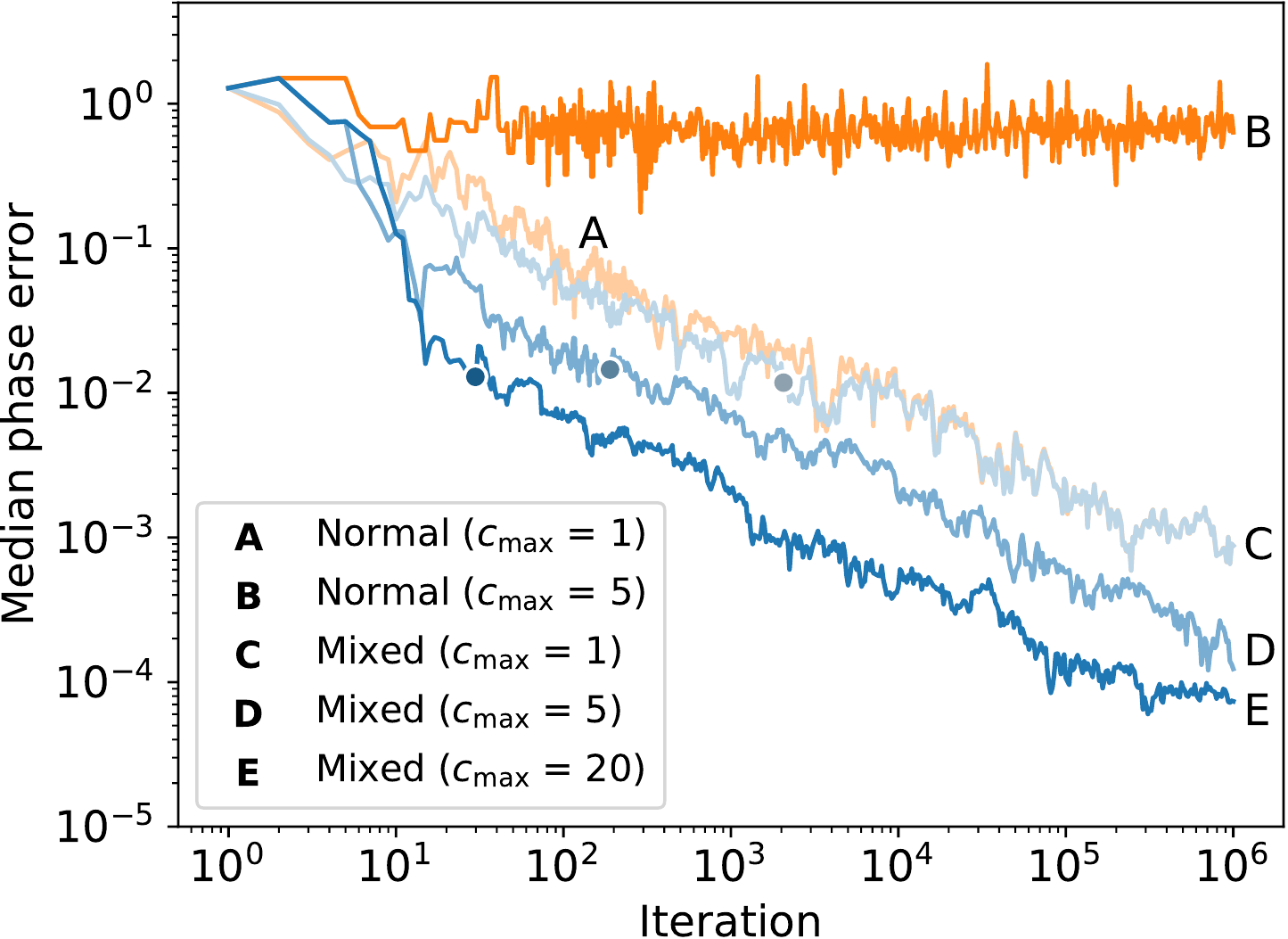} &
\includegraphics[width=0.305\textwidth]{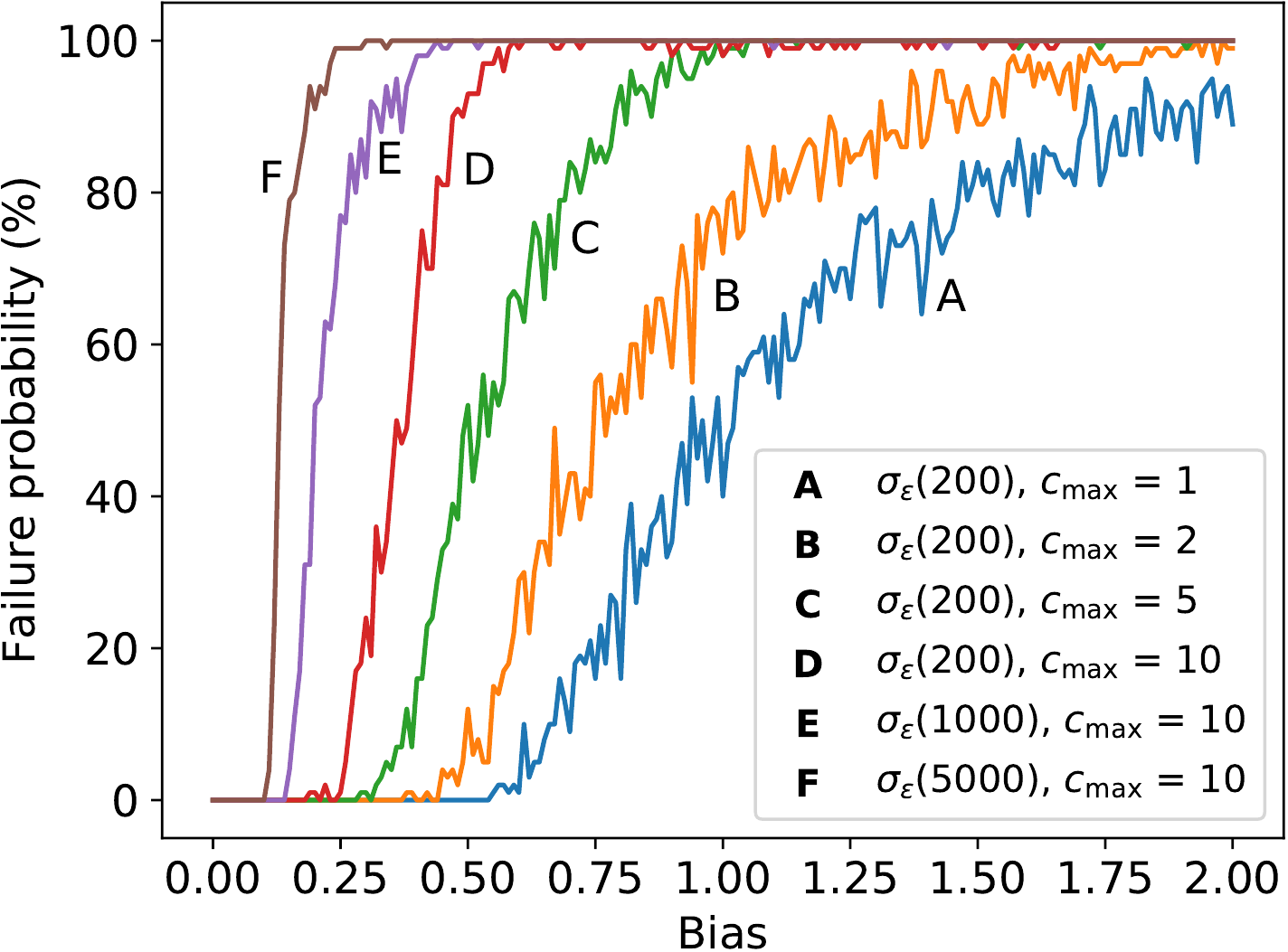} &
\includegraphics[width=0.305\textwidth]{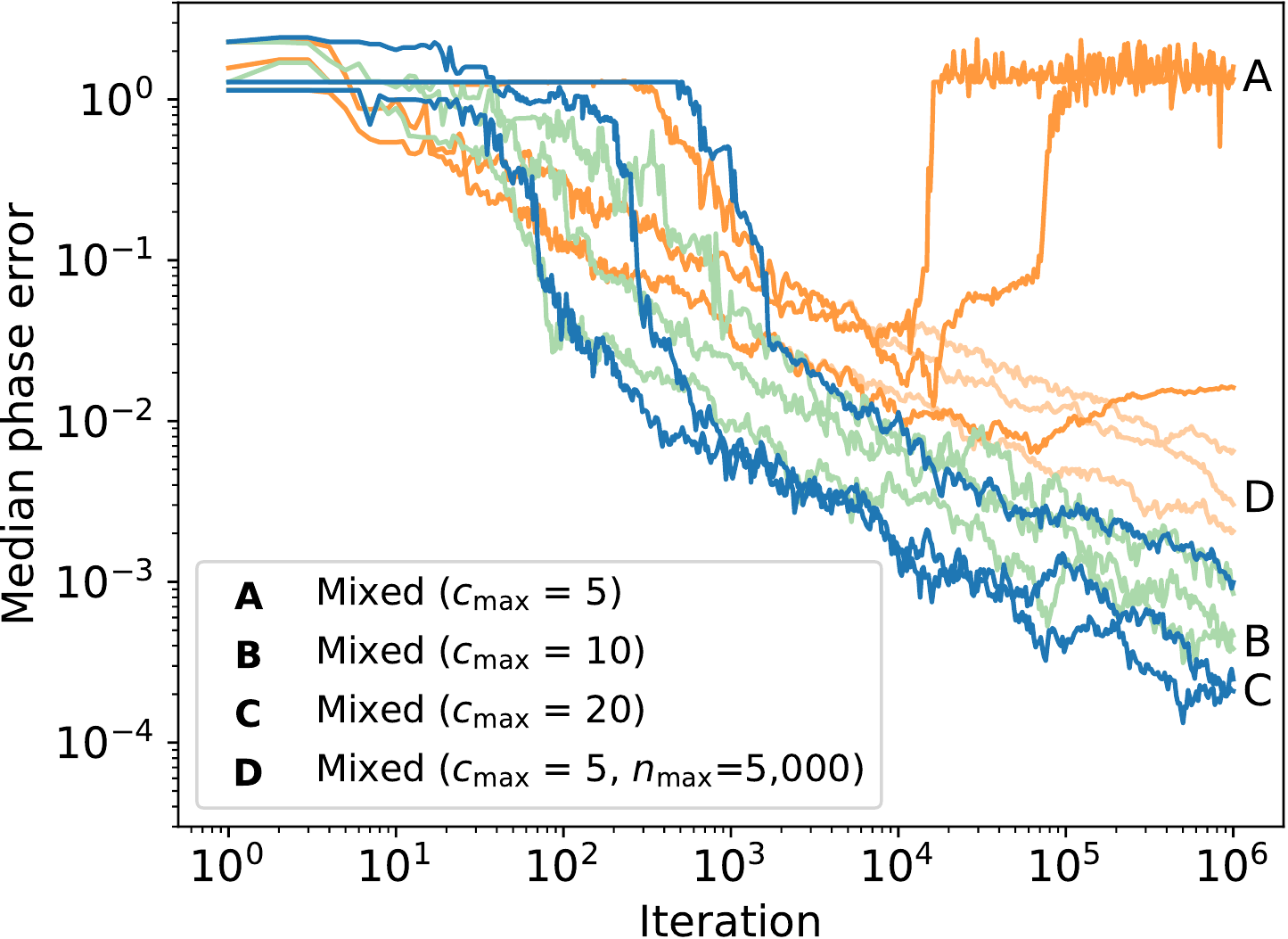} \\
({\bf{a}}) & ({\bf{b}}) & ({\bf{c}})\\[5pt]
\includegraphics[width=0.305\textwidth]{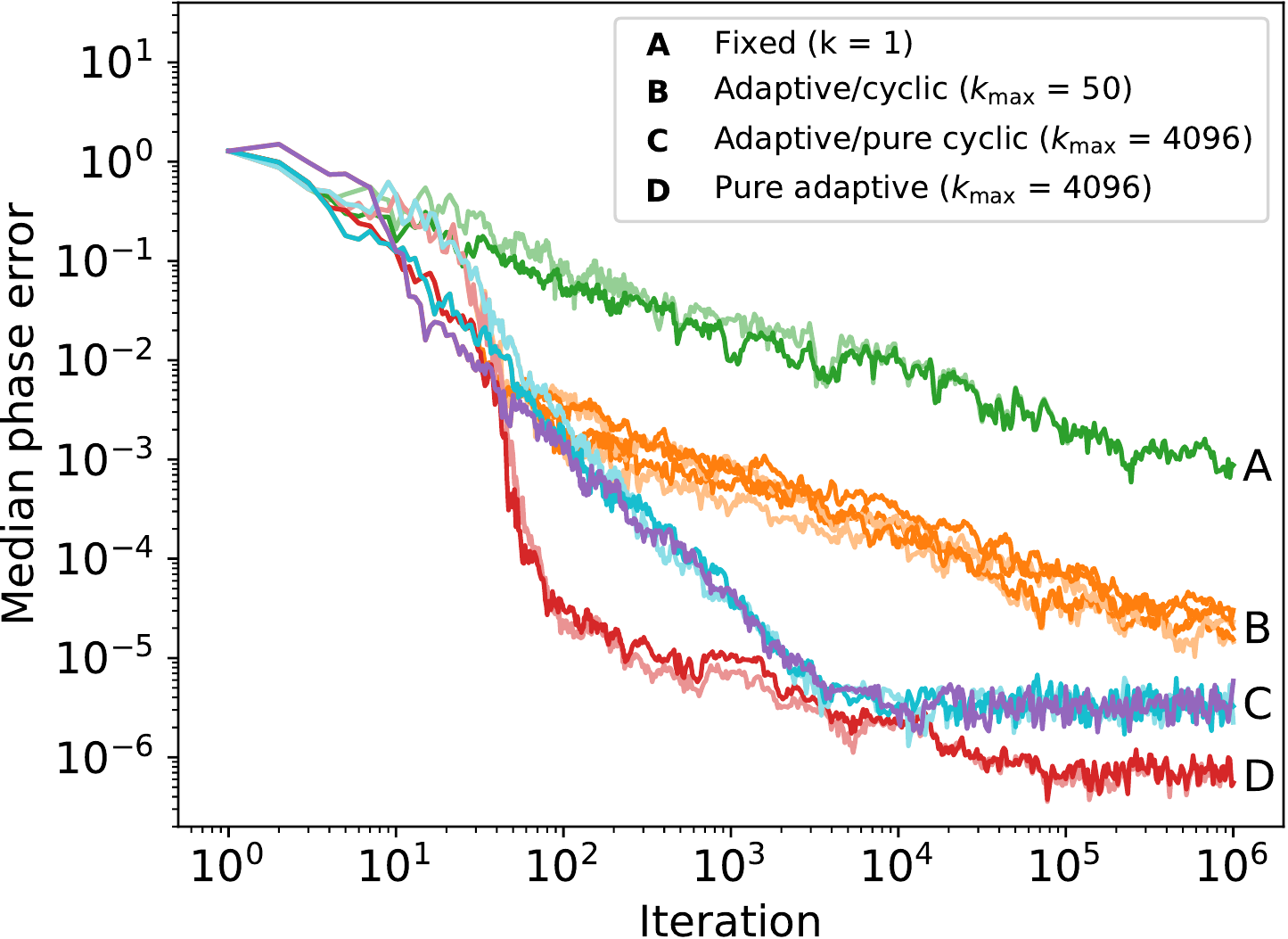} &
\includegraphics[width=0.305\textwidth]{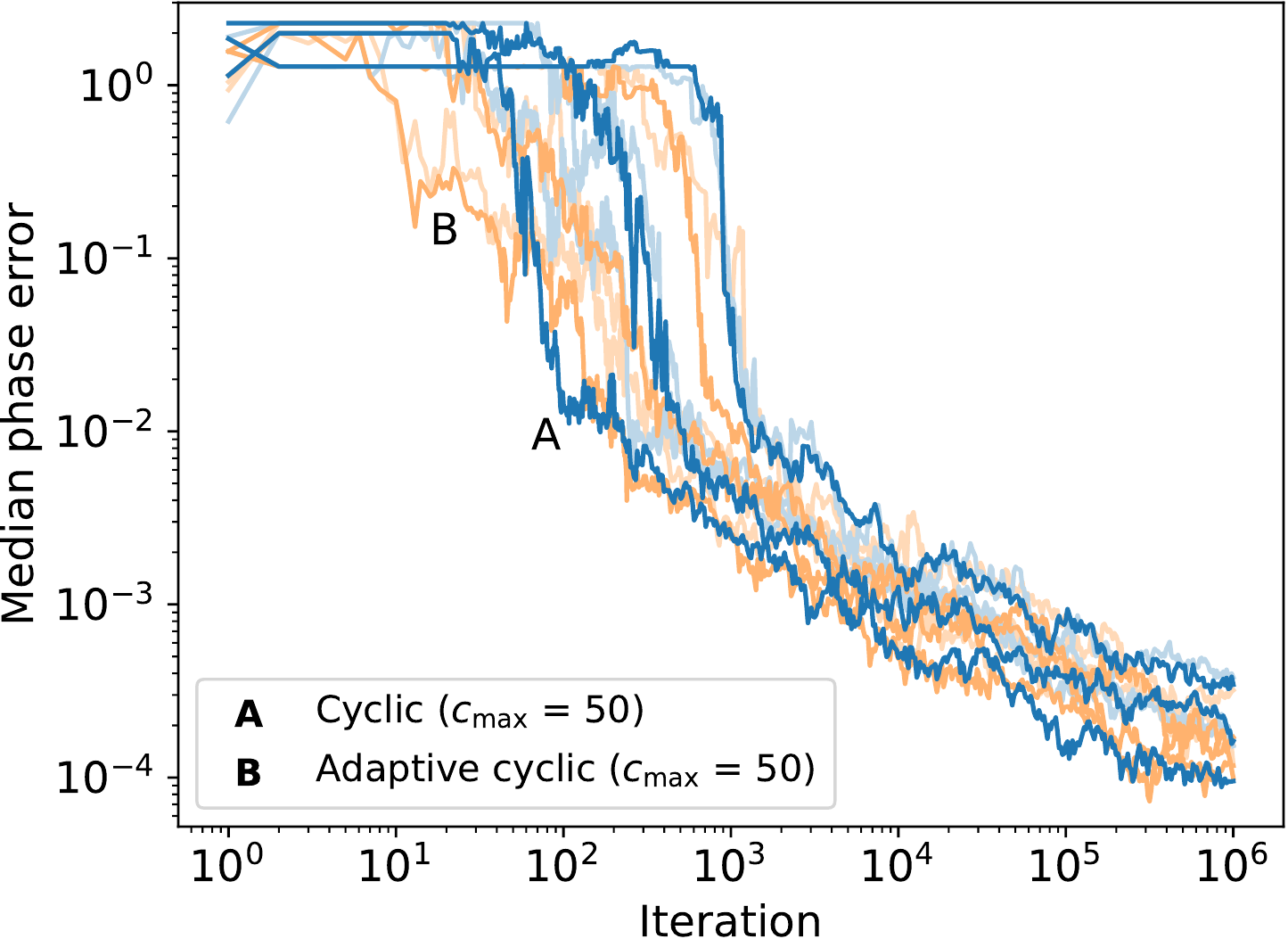} &
\includegraphics[width=0.305\textwidth]{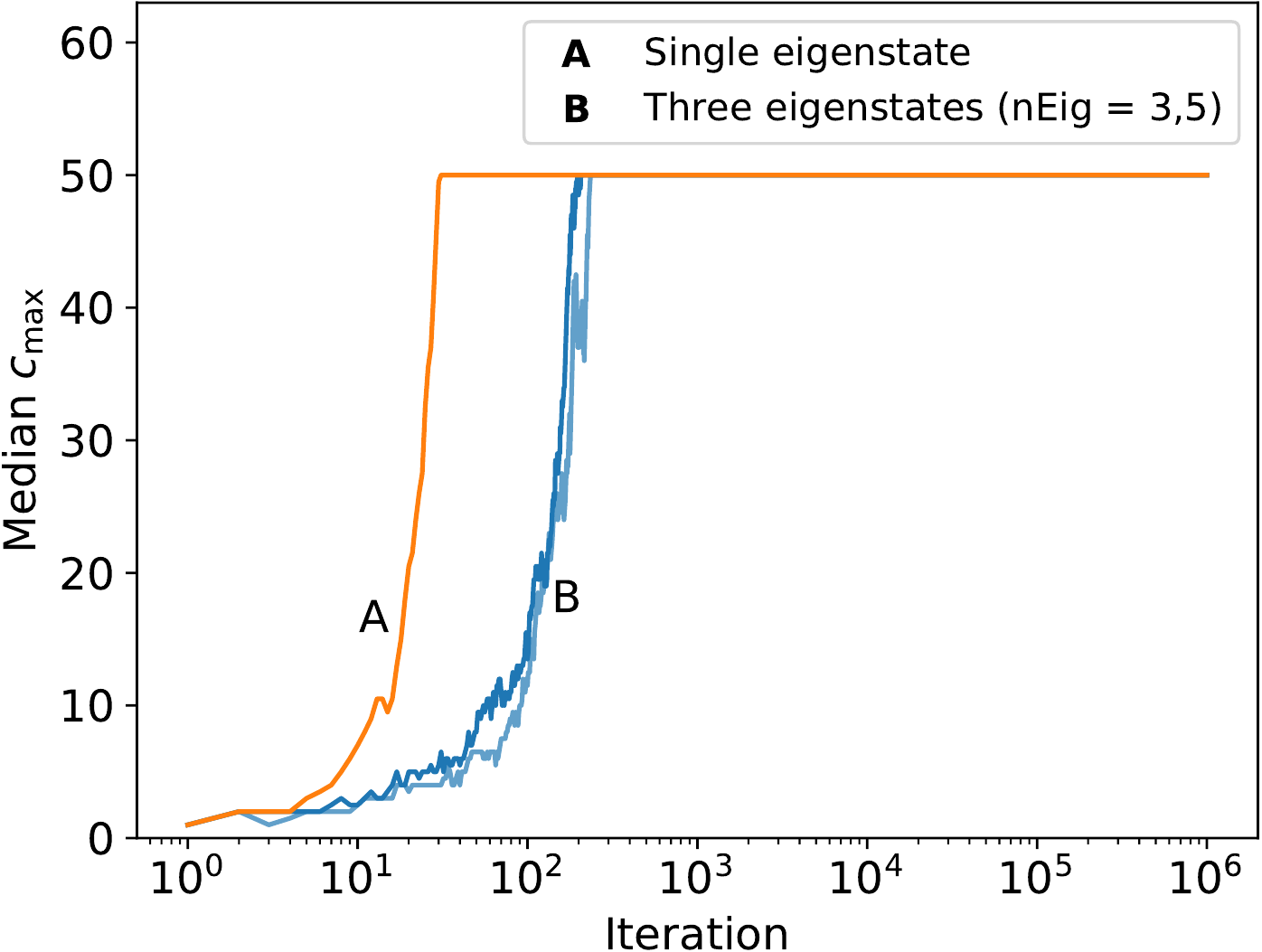} \\
({\bf{d}}) & ({\bf{e}}) & ({\bf{f}})
\end{tabular}
\caption{Plot (a) shows the convergence of the normal and mixed
  approaches in single-round experiments with cyclic $k$ in the range
  $1$ through $c_{\max}$. The three dots indicate the average
  iteration at which the mixed approach switches to the normal
  distribution. Plot (b) gives the probability of failure of the
  normal-based approach to converge as a function of deviation from
  the ideal phase for different initial standard deviations. Plot (d)
  shows the convergence of adaptive schemes in single-stage
  experiments with a single eigenphase using the normal (lighter
  shades) and mixed approach (darker shades); plot (e) shows the
  convergence of cyclic and adaptive cyclic schemes when applied to a
  problem with three eigenphases. The median $c_{\max}$ values
  resulting from the adaptive cyclic schemes from plots (d) and (e)
  are shown in plot (f). The curve with the lighter shade of blue
  corresponds to the setting where five distributions are used instead
  of three.}\label{Fig:ExperimentHybrid2}
\end{figure}

As an alternative to the three-round setup with fixed exponents
$\vec{k} = [1,2,5]$ used so far, we now consider the effect of using
single-round experiments in which the single $k$ value is either
chosen according to some fixed scheme or chosen adaptively based on
the estimated phase distributions. Using a fixed $k \neq 1$ was found
not to work, and we therefore follow~\cite{OBR2019TTa}, and consider
choosing $k$ cyclically over the values $\{1,2,\ldots,c_{\max}\}$ for
different values of $c_{\max}$. The resulting median phase error over
twenty problem instances with a single eigenphase is shown in
Figure~\ref{Fig:ExperimentHybrid2}(a). The performance of the normal
and mixed approaches are quite similar for $c_{\max} = 1$, which
coincides with keeping $k=1$ fixed. For $c_{\max}$ values equal to 2,
3, or 5, the normal-based approach consistently fails to converge to
the desired phase. The mixed approach, by contrast, shows rapid
convergence during the initial iterations and steady convergence for
subsequent iterations, even after switching to the normal
distribution.
The fact that the performance remains good must be due to the
initialization of the normal distribution with the partially converged
Fourier representation. To verify this, we study the convergence of
the normal approach when initialized with different normal
distributions. The standard deviation of all distributions is set to
the critical value $\sigma_{\epsilon}(200) \approx 0.023$, and the
mean is set to the correct phase plus or minus a bias term, for a
range of different bias values. We run 1,000 trials of the algorithm
for 100,000 iterations, using different cyclic schemes. The percentage
of trials that fail to converge to within 0.1 of the desired phase is
plotted in Figure~\ref{Fig:ExperimentHybrid2}(b) for each
setting. Note that those trials that do succeed generally continue to
converge to the desired phase as the number of iterations
increases. For each of the curves there is a clear transition between
bias values for which the method converges, and those for which the
algorithm fails to converge. As $c_{\max}$ increases, or the critical
sigma values decreases, this transition becomes sharper and starts at
smaller bias values. For the transition from a Fourier
representation to a normal distribution to succeed it is therefore
important that sufficient convergence has been achieved at the time of
the switch. If needed, this can be achieved by adjusting the scheme or
increasing the maximum number of Fourier coefficients.

Next, we apply the same schemes to the three-eigenstates problem
described in Section~\ref{Sec:ExperimentsNormalMixed}. When combined
with the normal-based approach, none of these schemes showed
successful convergence to the desired phases. Moreover, for
$c_{\max} = 1$ both the normal and mixed approaches failed to
converge. For other small values up to $c_{\max} = 5$, as seen in
Figure~\ref{Fig:ExperimentHybrid2}(c), the mixed approach initially
converged to the desired phases, but quickly diverged after switching
to the normal distribution. The plot also illustrates that we can
improve convergence by increasing $c_{\max}$ or by increasing the
number of terms used in the Fourier representation. Results obtained
when using five distributions to approximate the desired three are
similar to those shown and therefore omitted from the plot.

It is also possible to use an adaptive scheme in which the value of
$k_1$ is determined based on the current standard deviation. The
choice of $k_1 = \lceil 1.25/\sigma\rceil$ was proposed
in~\cite{WIE2016Ga} for single eigenstate experiments, and combined
with cyclic schemes for multiple eigenstate experiments
in~\cite{OBR2019TTa}. To evaluate the use of adaptive schemes in both
the normal-based and mixed approaches, we consider the quantity
\[
\bar{k} := \min\left\{\left\lceil \textstyle\sum_j
    \frac{1.25 w_j}{\sigma_j}\right\rceil,\ k_{\max}
\right\}.
\]
Based on this quantity we derive two schemes: a purely adaptive scheme
in which we simply set $k_1 = \bar{k}$, and an adaptive cyclic scheme
in which we choose $k$ cyclically with an adaptive maximum value
$c_{\max} = \bar{k}$. In Figure~\ref{Fig:ExperimentHybrid2}(d) we show
the performance of these schemes, along with fixed ($k=1$) and purely
cyclic schemes, on a single eigenstate problem. The adaptive and
cyclic schemes with $c_{\max} = 50$ have similar convergence and all
outperform the scheme with fixed $k_1$. Choosing $c_{\max} =4,096$
gives even faster convergence, especially so for the purely adaptive
scheme. The median phase error of the adaptive and purely cyclic
schemes steadily decreases until it comes to halt around iteration
$7,000$. Applying the same schemes to a problem with three
eigenphases, we find that none of the normal-based setups works. In
addition, the fixed and purely adaptive schemes, which previously
excelled, also failed to converge to the three desired values. Both
the purely cyclic and adaptive cyclic approaches converged to the
desired values, as shown in Figure~\ref{Fig:ExperimentHybrid2}(e). The
lighter-shade curves in the figure show that these methods continue to
work when using five rather than three distributions. The parameters
obtained using the adaptive cyclic approach are plotted in
Figure~\ref{Fig:ExperimentHybrid2}(f).

\subsection{Decoherence and read-out errors}

\begin{figure}[t!]
\centering
\begin{tabular}{ccc}
\includegraphics[width=0.305\textwidth]{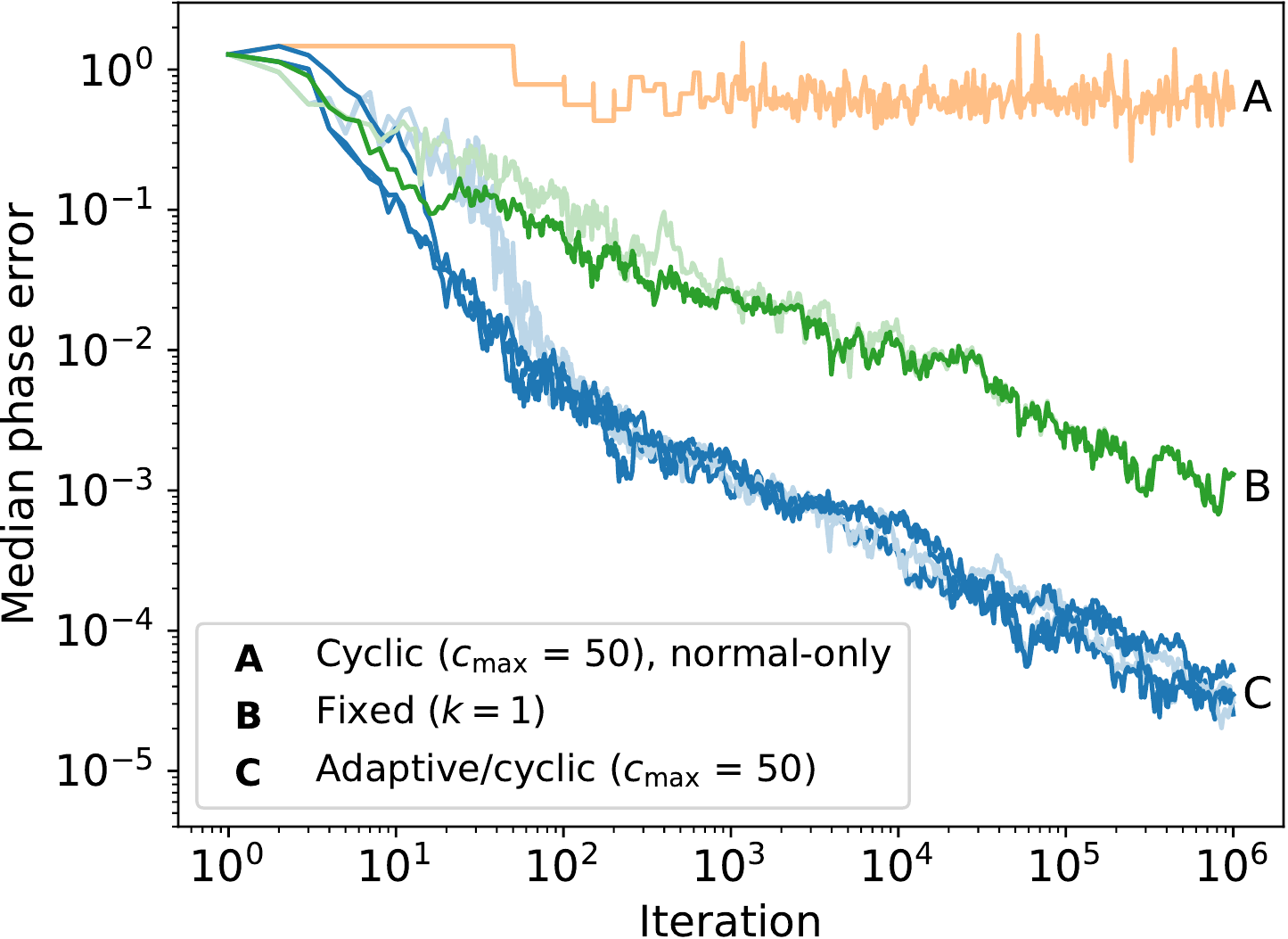} &
\multicolumn{2}{c}{\raisebox{6pt}{\fcolorbox{lightgray}{white}{\begin{minipage}[t]{0.610\textwidth}
\raisebox{9pt}{\includegraphics[width=\textwidth]{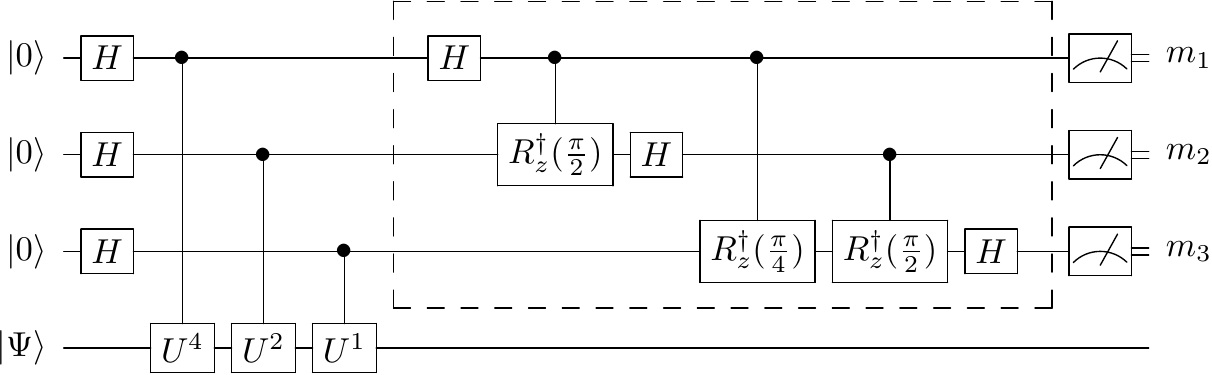}}
\end{minipage}}}}\\
({\bf{a}}) & \multicolumn{2}{c}{({\bf{c}})}\\[5pt]
\includegraphics[width=0.305\textwidth]{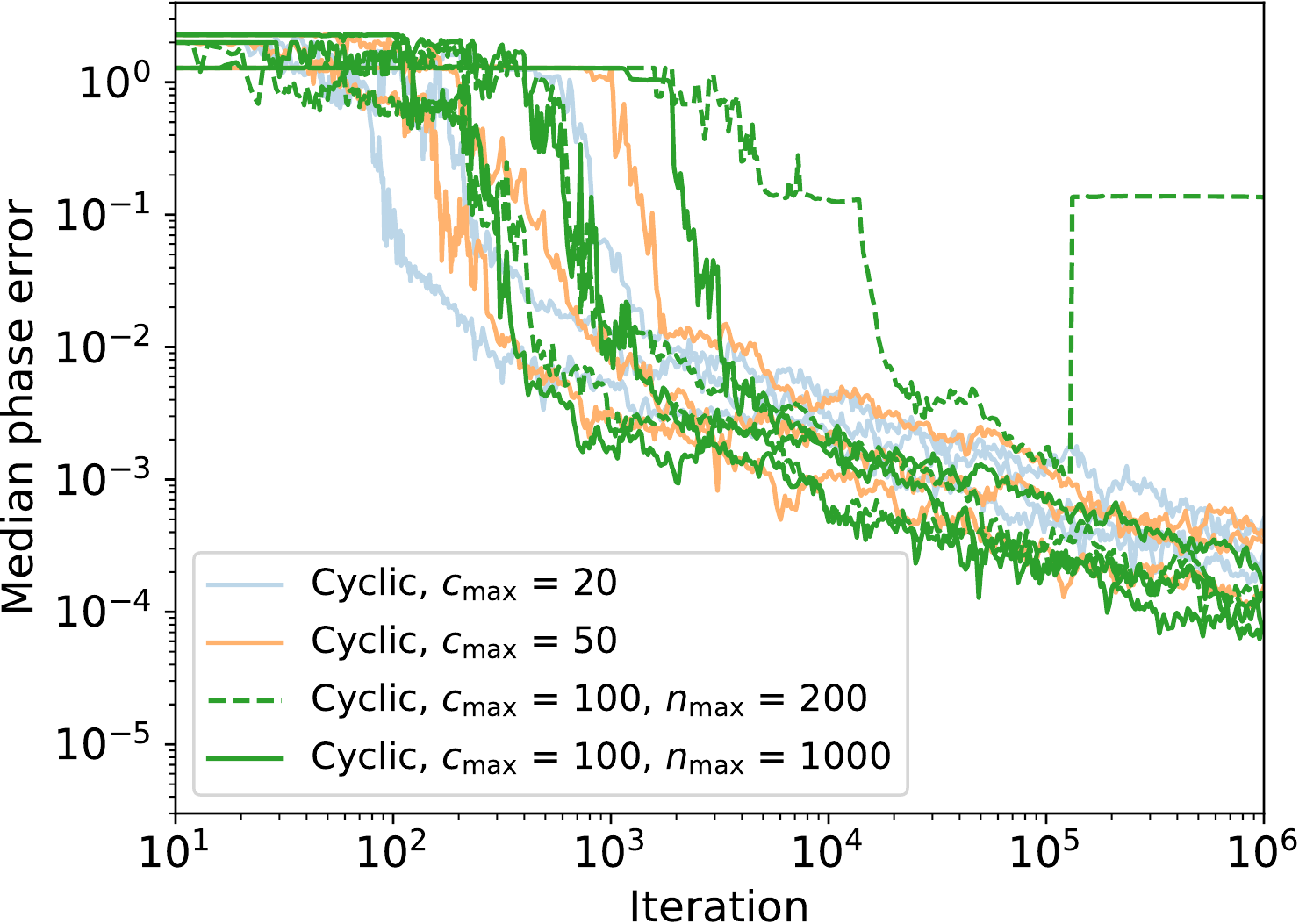} &
\includegraphics[width=0.305\textwidth]{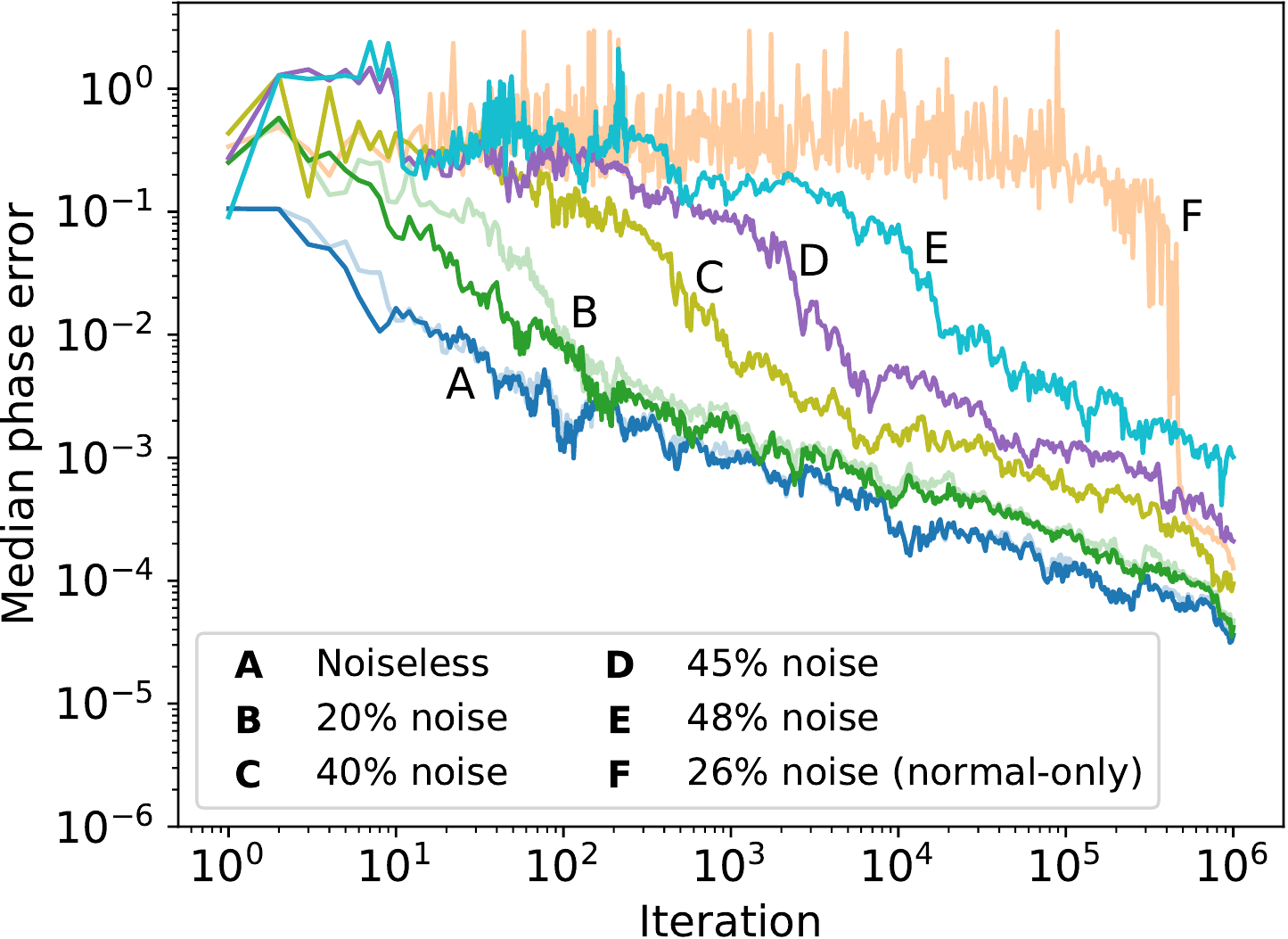} &
\includegraphics[width=0.305\textwidth]{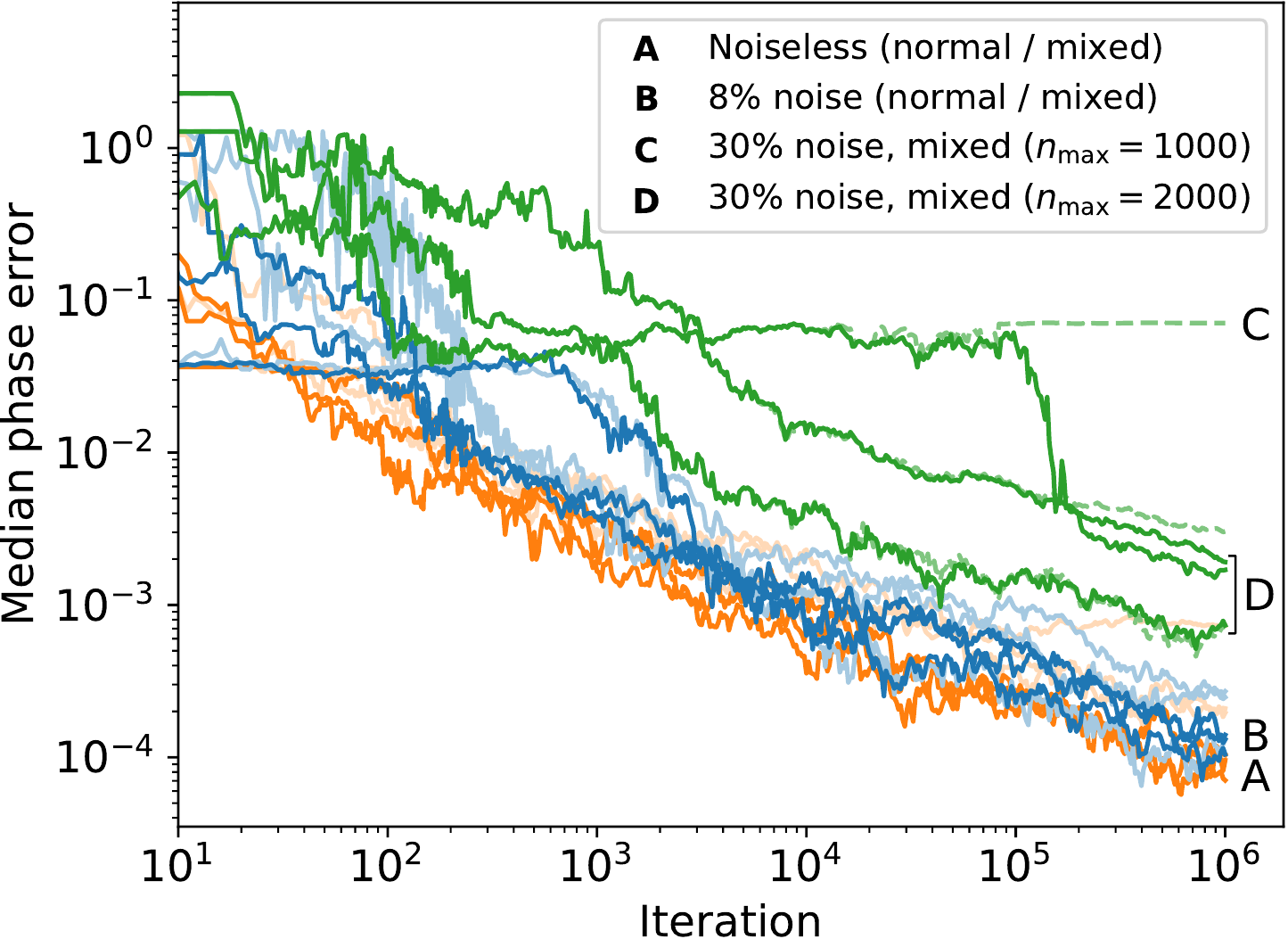} \\
({\bf{b}}) & ({\bf{d}}) & ({\bf{e}})
\end{tabular}
\caption{Convergence of the different Bayesian approaches with
  decoherence on problems with (a) a single eigenphase, and (b) three
  eigenphases. Plot (c) illustrates the circuit for quantum Fourier
  transformation with measurements on three qubits. The performance of
  Bayesian phase estimation subject to read-out errors is shown for
  (d) problems with a single eigenphase, and (e) problems with three
  eigenphases.}\label{Fig:Noise}
\end{figure}

In Section~\ref{Sec:Noise} we showed how different types of noise can
be included into the probability models. Here we study the performance
Bayesian phase estimation under decoherence and read-out errors.
Decoherence in single-round experiments has the effect of obtaining a
completely random measurement with probability
$1-e^{-k1/k_{\mathrm{err}}}$, where $k_1$ is the number of controlled
application of unitary $U$, and $k_{\mathrm{err}}$ is a system
dependent parameter, which we set to $100$ in our experiments. It can
be seen that the larger the value of $k_1$ the larger the probability
of obtaining a completely random measurement. Indeed, even for
$k_1 = 70$ this already occurs with a probability slightly exceeding
$0.5$, which clearly shows that we can no longer choose arbitrarily
large values for $k_1$. We therefore run adaptive and cyclic
experiments on a single eigenstate with $c_{\max}$ and $k_{\max}$
conservatively bounded by $50$. The resulting phase estimates along
with those obtained for a fixed $k_1 = 1$ are plotted in
Figure~\ref{Fig:Noise}(a). The cyclic and adaptive schemes have faster
initial convergence than the fixed scheme, and likewise for the mixed
versus the normal approach. The only configuration that failed was the
normal approach combined with a purely cyclic scheme.  When applied to
the problem with three eigenphases, we found that the purely adaptive
approach fails, even with $k_{\max}$ reduced to 10 or 20. The results
obtained using the mixed approach and the purely cyclic scheme, are
shown in Figure~\ref{Fig:Noise}(b). The results for the adaptive
cyclic scheme are similar aside from slightly faster initial
convergence and a somewhat slower convergence later. For the cyclic
scheme we observe that the convergence of the different phases begins
at later iterations when $c_{\max}$ is larger. In the case of
$c_{\max} = 100$ we see an abrupt degradation of the performance
around iteration $10^5$. As discussed in
Section~\ref{Sec:ExperimentsFourierRepr}, this happens when the
deviation to the correct phase is still too large at the time of
switching to the normal distribution. In this case we can avoid the
degradation by increasing the number of terms $n_{\max}$ in the
Fourier representation to 1000, as illustrated in
Figure~\ref{Fig:Noise}(b).

For our experiments with read-out errors, we apply Bayesian phase
estimation in combination with the quantum Fourier transform
(QFT). The QFT circuit, illustrated in Figure~\ref{Fig:Noise}(c) for
three measurement qubits, has long been used for non-iterative
estimation of a single eigenphase~\cite{ABR1999La,CLE1998EMMa}. Here
we show that the same circuit can be used to recover multiple
eigenphases even when the measurements are noisy. The circuit in
Figure~\ref{Fig:Noise}(c) is similar to the one given in
Figure~\ref{Fig:ExperimentCircuit}, although there are some important
differences. The first difference is that the exponents $\vec{k}$ are
fixed to successive powers of 2, starting at 1. The second difference
is that the phase-shift values $\vec{\beta}$ are no longer
predetermined but instead appear as conditional rotations. In our
example we use quantum-controlled gates, which means that we do not
know the effective value of the rotation until after measuring the
qubit, provided there are no read-out errors. An alternative
implementation where the rotations are classically controlled after
taking the measurements is also possible, but we do not consider it
here. Unlike our earlier experiment setup, we are now in a situation
where the effective $\vec{\beta}$ values depend on the quantum state
of the system. In order to deal with read-out errors we must therefore
distinguish between the actual state of the system, and the observed
measurements. As described in Section~\ref{Sec:Noise}, we determine
the measurement probability by first computing the probability of
ending up in each of the $2^n$ possible states, and multiplying this
by the probability of obtaining the observed measurement from the
given states. Although the computational complexity of this approach
grows exponential in the number qubits, it remains tractable when the
number of qubits is small. This is certainly the case for our
experiments, where we use a QFT circuit with measurements on five
qubits (note that the illustration in Figure~\ref{Fig:Noise}(c) uses
only three). The results obtained in this way on a single-eigenphase
problem with varying levels of read-out error are shown in
Figure~\ref{Fig:Noise}(d). The performance of the normal approach
gradually deteriorates as the noise level is increased, and the method
only just manages to converge when the read-out error reaches
26\%. Beyond that point the method does not converge to the desired
value within the given number of iterations. The mixed approach also
converges slower with increasing read-out errors, but performs well up
to a 48\% error rate. As seen from Figure~\ref{Fig:Noise}(e), the
normal and mixed approach both converge for the problem with three
eigenphases with read-out error rates up to 8\%. When the noise
reaches 10\% or above, the normal-based approach fails to find the
desired phases. For the mixed approach this happens at noise levels of
around 30\%.

\subsection{Additional phases}

\begin{figure}
\centering
\begin{tabular}{ccc}
\includegraphics[width=0.305\textwidth]{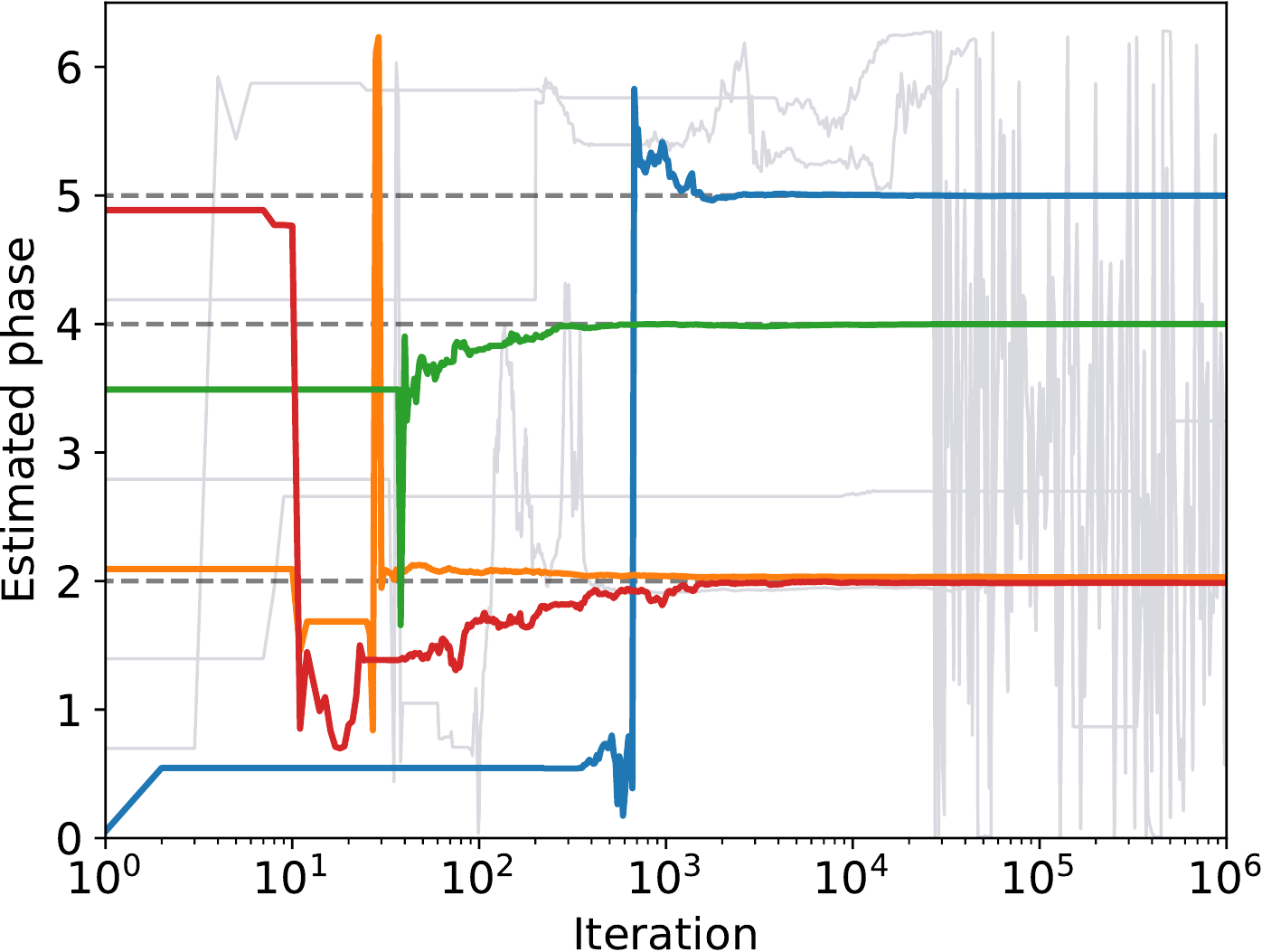} &
\includegraphics[width=0.305\textwidth]{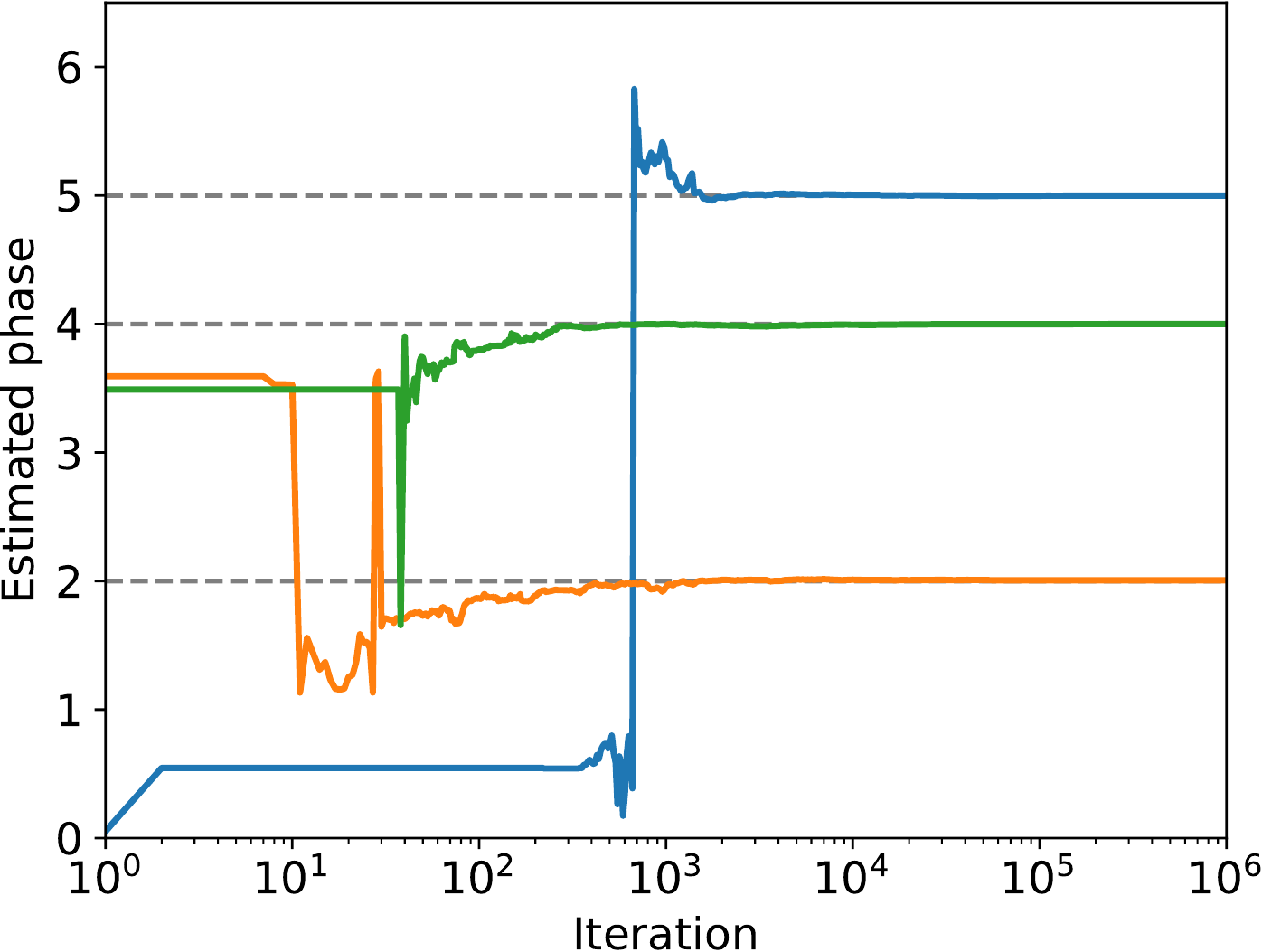}&
\includegraphics[width=0.305\textwidth]{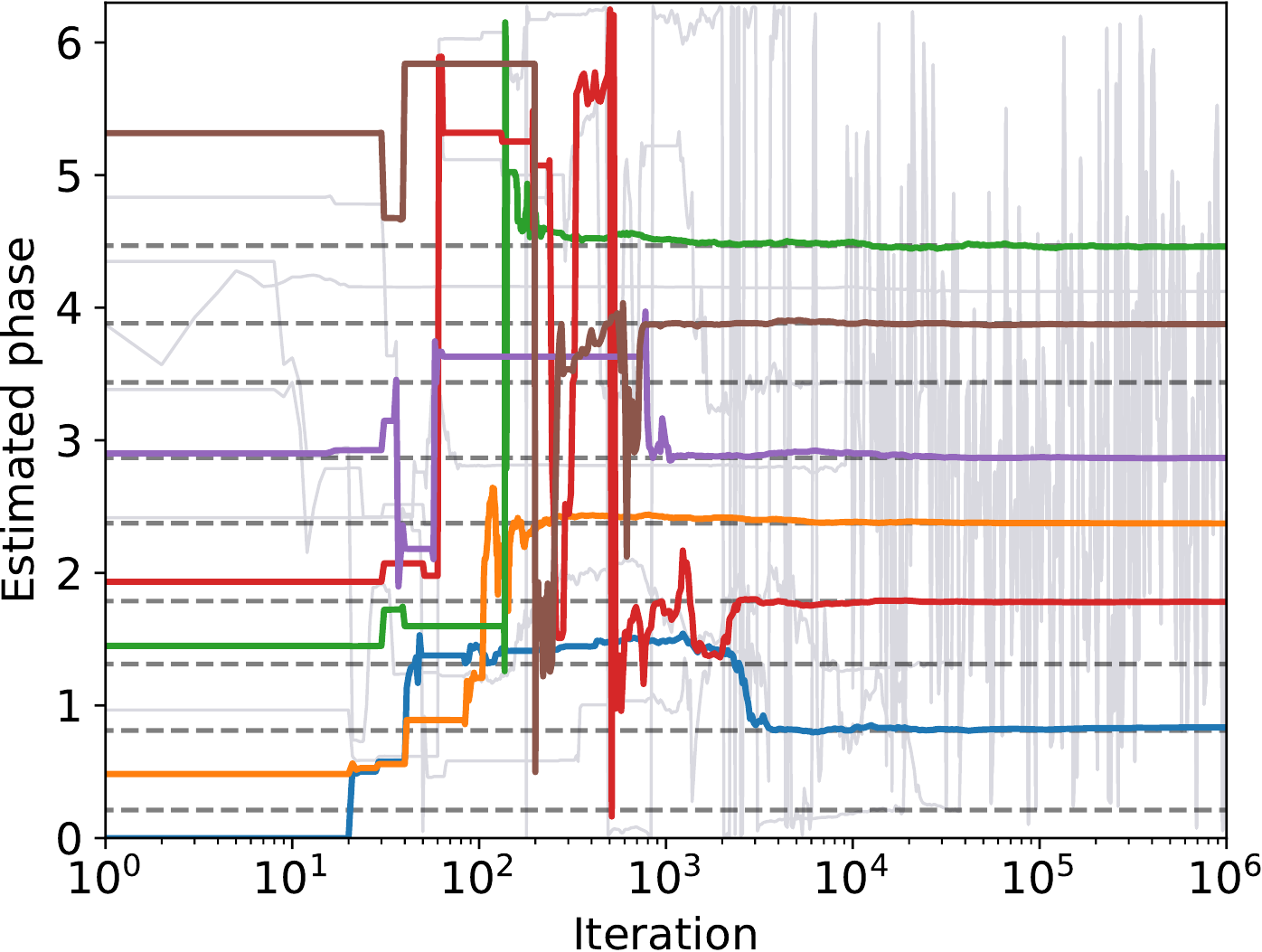}\\
({\bf{a}}) & ({\bf{b}}) & ({\bf{c}})
\end{tabular}
\caption{Plots of (a) the trajectories of eight phase estimates in the
  case of three dominant phases with values indicated by the
  horizontal dotted lines. Phase estimates with low final weight are
  shown in light gray; (b) phase estimates after filtering out low
  weight and highly oscillatory estimates and combining bundled
  estimates; and (c) phase estimates for a problem instance with nine
  phases.}\label{Fig:Spurious}
\end{figure}

In practical settings, it can be expected that the initial state is a
superposition of not only the eigenstates of interest but also one or
more spurious eigenstates of small weight. To get a better
understanding of the effect these eigenstates have on phase
estimation, we revisit our earlier experiments consisting of three
main phases, but this time include additional eigenstates. In a series
of problem instances, we vary the number of additional phases while
keeping their total weight to $0.1$, and scale down the weights of the
original three phases by a factor of $0.9$, yielding weighs
$[0.45, 0.27, 0.18]$.  For phase estimation, we use the mixed approach
with depths $k$ chosen according to the cyclic scheme with
$c_{\max} = 20$. We provide the algorithm with extra phase parameters,
which are needed absorb the spurious phases and also reflect the fact
that the actual number of eigenstates in the initial phase is
typically unknown in practice. In Figure~\ref{Fig:Spurious}(a) we show
the trajectory of eight phase estimates in the case of two spurious
phases. From the given trajectories we would like to extract the three
dominant phases, indicated by the horizontal dashed lines. There are
three simple techniques we use to do so. First, we note that two of
the estimated phases in this example have zero weight, and these are
therefore easily filtered out. More generally, we can filter out
phases with weights below some selected threshold. Second, we see that
several of the phases exhibit strong oscillations. Clearly, these
phases cannot provide an accurate estimate, and we therefore filter
them out based on the total variation of the estimated angle, that is,
the sum of absolute differences between successive estimates, over the
last, say 50, recorded values. Third, we see that multiple estimated
phases can converge towards the same phase; $\phi_1=2$ in this
case. We found that such estimates tend to repel each other, leading
to the situation where none of them is accurate. To correct this, we
detect bundles of similar estimates (phases deviating at most $\tau$
degrees from the bundle center) and replace them by their weighted
average. Applying these three correction steps for the current example
using $\tau=5$ we obtain the phase estimates shown in
Figure~\ref{Fig:Spurious}(b).  Next, we generate 100 problem instances
each for $2$, $5$, $10$, and $20$ spurious phases with weights and
phases sampled uniformly at random. We normalize the total weight of
the spurious phases such that they add up to 0.1 and apply Bayesian
phase estimation followed by the filtering process described
above. For each problem instance we record the weight and phase
differences whenever exactly three phases are recovered, otherwise we
mark the problem instance as failed. We run the algorithm with three
plus a given number of additional phase parameters for threshold
values $\tau \in \{1,3,5\}$, and summarize the results in
Table~\ref{Table:Spurious}. For the successful cases with $\tau=3$ we
see that the phase error is small in all but one setting, indicating
that whenever the algorithm converged, it did so to the correct
phases. The fact that the maximum phase error decreases as the number
of spurious phases increases is likely due to the decrease of the
individual weights. The worst-case estimation error in the weights is
worse than that of the phases, but overall the median error is very
small. The number of failed cases decreases significantly with larger
values of $\tau$, which is possible here because the actual phases are well
spread out. Finally, we note that the method seems fairly insensitive
to the selected number of additional phase parameters.

\begin{table}[t]
\centering
\begin{tabular}{|lllll|rrrr|}
\hline
spurious & additional & \multicolumn{3}{c}{\# failed} & \multicolumn{2}{c}{phase error} & \multicolumn{2}{c|}{weight error}\\
phases & parameters & \multicolumn{3}{c}{instances} & maximum & median & maximum & median\\
\hline
2 & 1 & 35&34&26 & 0.02072 & 0.00090 & 0.12520 & 0.00458\\
2 & 2 & 32&25&20 & 0.01896 & 0.00086 & 0.08319 & 0.00373\\
2 & 3 & 32&21&19 & 0.01968 & 0.00085 & 0.08277 & 0.00339\\
2 & 4 & 35&24&21 & 0.01968 & 0.00081 & 0.08212 & 0.00308\\
2 & 5 & 33&25&24 & 0.02093 & 0.00083 & 0.08175 & 0.00293\\
\hline
5 & 1 & 19&18&9 & 0.01100 & 0.00077 & 0.06490 & 0.00310\\
5 & 2 & 19&7&2 & 0.01119 & 0.00076 & 0.06445 & 0.00279\\
5 & 3 & 15&8&1 & 0.01087 & 0.00079 & 0.06374 & 0.00284\\
5 & 4 & 15&3&0 & 0.01083 & 0.00081 & 0.06421 & 0.00261\\
5 & 5 & 22&4&0 & 0.01044 & 0.00077 & 0.06395 & 0.00285\\
\hline
10 & 1 & 18&18&8 & 0.01260 & 0.00085 & 0.08789 & 0.00267\\
10 & 2 & 26&15&1 & 0.01232 & 0.00085 & 0.05039 & 0.00276\\
10 & 3 & 22&9&1 & 0.01276 & 0.00085 & 0.03364 & 0.00272\\
10 & 4 & 18&4&0 & 0.01225 & 0.00078 & 0.03106 & 0.00262\\
10 & 5 & 16&1&0 & 0.01174 & 0.00086 & 0.05797 & 0.00272\\
\hline
20 & 1 & 25&25&8 & 1.95686 & 0.00082 & 0.14688 & 0.00315\\
20 & 2 & 22&16&0 & 0.00644 & 0.00082 & 0.03081 & 0.00262\\
20 & 3 & 20&8&0 & 0.00607 & 0.00076 & 0.02692 & 0.00263\\
20 & 4 & 18&8&0 & 0.00645 & 0.00076 & 0.02183 & 0.00248\\
20 & 5 & 28&8&0 & 0.00668 & 0.00076 & 0.02404 & 0.00269\\
\hline
\end{tabular}
\caption{Performance of the Bayesian phase estimation algorithm with
  filtering of the results for 100 problem instances with three base
  phases and varying numbers of spurious phases. The algorithm uses
  three plus the given number of additional phase
  parameters. Clustering of phase estimates is done with threshold
  values $\tau$ equal to 1, 3, and 5 respectively, and the
  corresponding number of problem instances that failed to return
  three phases are listed, from left to right, in the failed instances
  column. For threshold parameter $\tau=3$
  we show the maximum  and median phase and weight errors over all
  successful problem instances.}\label{Table:Spurious}
\end{table}

\subsection{Scalability}

We now move beyond the three phases considered so far and look at the
performance of Bayesian phase estimation on problems with up to twelve
phases. For this we consider a somewhat idealized setting where the
unknown phases are initialized at a grid of angles with $\pi/6$
increments starting at $\pi/12$. We randomly perturb these values by
adding angles, in radians, sampled uniformly at random from the
interval $[-0.05,0.05]$. The weights are sampled uniformly at random
from $[\half,1]$ to ensure all weights are similar in magnitude, and
are then normalized to sum up to one. We generate 100 instances each
for problems with the number of phases ranging from one to twelve and
apply the filtering procedure described in the previous section to
obtain the final estimates. For this set of test problems we do not
include any spurious phases. We focus on recovery of the phases and
declare a problem instance to be successful if it recovers all
ground-truth phases with a deviation of at most 0.005 radians, and it
does not include any other phases. Table~\ref{Table:Multiple}
summarizes the success counts for filtering with threshold values
$\tau \in \{0.01,1,3,5\}$ and a varying number of additional phase
parameters.  Increasing $\tau$ results in higher success rates, but as
mentioned above, this comes at the cost of resolution and can be done
only when the threshold is less than half the minimum distance between
the phases. For those problem instances that failed to recover all
phases, we found that the estimated values were often accurate, but
that some of the phases were not included, as illustrated in
Figure~\ref{Fig:Spurious}(c) for a problem instance with nine phases.
Further simulations would be needed to see how the algorithm deals
with phases that are clustered together more closely or exhibit a
larger range of weights.

\begin{table}
\centering
\begin{tabular}{|l|l|llllllllllll|}
\hline
phase&&\multicolumn{12}{|c|}{$n$}\\
parameters & $\tau$&  1&  2&  3&  4&  5&  6&  7&  8&  9& 10& 11& 12\\
\hline
$n$+0 & 0.01& 100&  96&  90&  85&  78&  66&  70&  39&   9&   0&   0&   0\\
$n$+1 & &  97&  87&  89&  83&  89&  85&  77&  52&  11&   0&   0&   0\\
$n$+2 & &  98&  85&  82&  89&  85&  87&  84&  61&  17&   0&   0&   0\\
$n$+3 & &  99&  86&  74&  81&  81&  89&  81&  64&  13&   3&   0&   0\\
\hline
$n$+0 & 1& 100&  96&  90&  85&  78&  66&  70&  39&   9&   0&   0&   0\\
$n$+1 & & 100&  91&  93&  88&  90&  85&  77&  52&  11&   0&   0&   0\\
$n$+2 & & 100&  91&  84&  91&  87&  87&  84&  61&  17&   0&   0&   0\\
$n$+3 & & 100&  90&  82&  83&  81&  89&  81&  64&  13&   3&   0&   0\\
\hline
$n$+0 & 3& 100&  96&  90&  85&  78&  66&  70&  39&   9&   0&   0&   0\\
$n$+1 & & 100& 100&  99&  97&  94&  90&  79&  52&  11&   0&   0&   0\\
$n$+2 & & 100& 100&  99&  99&  98&  91&  87&  62&  17&   0&   0&   0\\
$n$+3 & & 100& 100& 100&  99&  98&  95&  82&  64&  13&   3&   0&   0\\
\hline
$n$+0 & 5& 100&  96&  90&  85&  78&  66&  70&  39&   9&   0&   0&   0\\
$n$+1 & & 100& 100& 100&  99&  98&  92&  79&  52&  11&   0&   0&   0\\
$n$+2 & & 100& 100& 100& 100&  99&  92&  87&  62&  17&   0&   0&   0\\
$n$+3 & & 100& 100& 100&  99&  99&  95&  82&  64&  13&   3&   0&   0\\
\hline
\end{tabular}
\caption{Number of successful phase estimates for 100 problem
  instances with $n$ phases, obtained using Bayesian phase estimation
  using the given number of phase parameters and post processing with
  bundle size $\tau$ (in degrees).}\label{Table:Multiple}
\end{table}

\section{Conclusions}\label{Sec:Discussion}

In this work we have analyzed the performance of Bayesian phase
estimation for different representations of the prior
distributions. As a first representation we use a normal distribution,
which was used earlier in~\cite{WIE2016Ga}. We show that updates to
the distributions can be evaluated analytically in noiseless settings,
as well as in settings with several types of commonly encountered
noise. The normal-based approach is fast, but performs poorly when
multiple eigenphases are present, certainly in experiments with only a
single measurement round. In these cases the distributions tend to
collapse into a single distribution, which is then impossible to
untangle since the updates will essentially be identical. As a second
representation, we consider the truncated Fourier series
representation used in~\cite{OBR2019TTa}. For noiseless problems, as
well as for problems with many different types of noise, the Fourier
series is ideal in that it captures exactly the desired
distributions. However, the number of coefficients required to
maintain exact representations keeps growing with each additional
round of measurement. In order for the method to remain
computationally attractive, it is therefore necessary to truncate the
Fourier series. We show that this truncation eventually causes the
distribution to become unstable, thereby limiting the accuracy that
can be achieved using a fixed number of terms. To combine the
advantages of both approaches, we propose a mixed approach in which
the distributions are initially represented as truncated Fourier
series. When successive updates cause the standard deviation of a
distribution to fall below a suitably chosen threshold, we change the
representation to a normal distribution. We show that the proposed
mixed approach performs well with both static and adaptively chosen
values of $k$, and that the performance remains stable when
decoherence or read-out errors are present. Finally, we show how the
Bayesian approach can be combined with the quantum Fourier
transformation, which is traditionally used for phase
estimation. Measurement errors can be dealt with successfully in this
setting but the current method requires the evaluation of the
probability distributions for all possible states. Reduction of this
complexity is an interesting topic for future work.
In terms of scalability of the method we conclude that Bayesian phase
estimation is especially well suited for problems where the state is a
superposition of a small number of dominant eigenstates and
possibly many spurious eigenstates with a much smaller weight.

\section*{Acknowledgments}

The author would like to thank Antonio C\'{o}rcoles and Maika Takita
for useful discussions, and the referees for feedback that helped
improve the paper.

\clearpage
\bibliography{bibliography}

\begin{thebibliography}{19}
\providecommand{\natexlab}[1]{#1}
\providecommand{\url}[1]{\texttt{#1}}
\expandafter\ifx\csname urlstyle\endcsname\relax
  \providecommand{\doi}[1]{doi: #1}\else
  \providecommand{\doi}{doi: \begingroup \urlstyle{rm}\Url}\fi

\bibitem[Abrams and Lloyd(1999)]{ABR1999La}
Daniel~S. Abrams and Seth Lloyd.
\newblock Quantum algorithm providing exponential speed increase for finding
  eigenvalues and eigenvectors.
\newblock \emph{Physical Review Letters}, 83\penalty0 (24):\penalty0
  5162--5165, 1999.
\newblock \doi{10.1103/PhysRevLett.83.5162}.

\bibitem[Aspuru-Guzik et~al.(2005)Aspuru-Guzik, Dutoi, Love, and
  Head-Gordon]{ASP2005DLHa}
Al{\'a}n Aspuru-Guzik, Anthony~D. Dutoi, Peter~J. Love, and Martin Head-Gordon.
\newblock Simulated quantum computation of molecular energies.
\newblock \emph{Science}, 309\penalty0 (5741):\penalty0 1704--1707, 2005.
\newblock \doi{10.1126/science.1113479}.

\bibitem[Berry et~al.(2009)Berry, Higgins, Bartlett, Mitchell, Pryde, and
  Wiseman]{BER2009HBMa}
Dominic~W. Berry, Brendon~L. Higgins, Stephen~D. Bartlett, Morgan~W. Mitchell,
  Geoff~J. Pryde, and Howard~M. Wiseman.
\newblock How to perform the most accurate possible phase measurements.
\newblock \emph{Physical Review A}, 80\penalty0 (5):\penalty0 052114, 2009.
\newblock \doi{10.1103/PhysRevA.80.052114}.

\bibitem[Bertsekas(1999)]{BER1999a}
Dimitri~P. Bertsekas.
\newblock \emph{Nonlinear Programming}.
\newblock Athena, 2nd edition, 1999.

\bibitem[Bryc(1995)]{BRY1995a}
Wlodzimierz Bryc.
\newblock \emph{The Normal Distribution: Characterizations with Applications}.
\newblock Springer-Verlag, 1995.
\newblock \doi{10.1007/978-1-4612-2560-7}.

\bibitem[Cleve et~al.(1998)Cleve, Ekert, Macchiavello, and Mosca]{CLE1998EMMa}
Richard Cleve, Arthur Ekert, Chiara Macchiavello, and Michele Mosca.
\newblock Quantum algorithms revisited.
\newblock \emph{Proceedings of the Royal Society A}, 454\penalty0
  (1969):\penalty0 339--354, 1998.
\newblock \doi{10.1098/rspa.1998.0164}.

\bibitem[Condat(2016)]{CON2016a}
Laurent Condat.
\newblock Fast projection onto the simplex and the $\ell_1$ ball.
\newblock \emph{Mathematical Programming, Series A}, 158\penalty0
  (1--2):\penalty0 575--585, 2016.
\newblock \doi{10.1007/s10107-015-0946-6}.

\bibitem[Grafakos(2008)]{GRA2008a}
Loukas Grafakos.
\newblock \emph{Classical {F}ourier analysis}.
\newblock Springer, 3rd edition, 2008.
\newblock \doi{10.1007/978-1-4939-1194-3}.

\bibitem[Hayes and Berry(2014)]{HAY2014Ba}
Alexander J.~F. Hayes and Dominic~W. Berry.
\newblock Swarm optimization for adaptive phase measurement with low
  visibility.
\newblock \emph{Physical Review A}, 89\penalty0 (1):\penalty0 013838, 2014.
\newblock \doi{10.1103/PhysRevA.89.013838}.

\bibitem[Kimmel et~al.(2015)Kimmel, Low, and Yoder]{KIM2015LYa}
Shelby Kimmel, Guang~Hao Low, and Theodore~J. Yoder.
\newblock Robust calibration of a universal single-qubit gate set via robust
  phase estimation.
\newblock \emph{Physical Review A}, 92\penalty0 (6):\penalty0 062315, 2015.
\newblock \doi{10.1103/PhysRevA.92.062315}.

\bibitem[Kitaev(1995)]{KIT1995a}
Alexei~{Yu}. Kitaev.
\newblock Quantum measurements and the {A}belian stabilizer problem.
\newblock arXiv preprint quant-ph/9511026, 1995.
\newblock URL \url{https://arxiv.org/abs/quant-ph/9511026}.
\newblock (See also Electronic Colloquium on Computational Complexity,
  TR96-003, 1996).

\bibitem[Nocedal and Wright(2006)]{NOC2006Wa}
Jorge Nocedal and Stephen~J. Wright.
\newblock \emph{Numerical Optimization}.
\newblock Springer series in operations research and financial engineering.
  Springer, second edition, 2006.
\newblock \doi{10.1007/978-0-387-40065-5}.

\bibitem[O'Brien et~al.(2019)O'Brien, Tarasinski, and Terhal]{OBR2019TTa}
Thomas~E. O'Brien, Brian Tarasinski, and Barbara~M. Terhal.
\newblock Quantum phase estimation of multiple eigenvalues for small-scale
  (noisy) experiments.
\newblock \emph{New Journal of Physics}, 21:\penalty0 023022, 2019.
\newblock \doi{10.1088/1367-2630/aafb8e}.

\bibitem[Roggero(2020)]{ROG2020a}
Alessandro Roggero.
\newblock Spectral-density estimation with the {G}aussian integral transform.
\newblock \emph{Physical Review A}, 102:\penalty0 022409, Aug 2020.
\newblock \doi{10.1103/PhysRevA.102.022409}.

\bibitem[Shor(1997)]{SHO1997a}
Peter~W. Shor.
\newblock Polynomial-time algorithms for prime factorization and discrete
  logarithms on a quantum computer.
\newblock \emph{SIAM Journal on Computing}, 26\penalty0 (5):\penalty0
  1484--1509, 1997.
\newblock \doi{10.1137/S0097539795293172}.

\bibitem[Somma(2019)]{SOM2019a}
Rolando~D. Somma.
\newblock Quantum eigenvalue estimation via time series analysis.
\newblock \emph{New Journal of Physics}, 21:\penalty0 123025, 2019.
\newblock \doi{10.1088/1367-2630/ab5c60}.

\bibitem[Svore et~al.(2014)Svore, Hastings, and Freedman]{SVO2014HFa}
Krysta~M. Svore, Matthew~B. Hastings, and Michael Freedman.
\newblock Faster phase estimation.
\newblock \emph{Quantum Information \& Computation}, 14\penalty0
  (3--4):\penalty0 306--328, March 2014.

\bibitem[Temme et~al.(2011)Temme, Osborne, Vollbrecht, Poulin, and
  Verstraete]{TEM2011OVPa}
Kristan Temme, Tobias~J. Osborne, Karl~G. Vollbrecht, David Poulin, and Frank
  Verstraete.
\newblock Quantum {M}etropolis sampling.
\newblock \emph{Nature}, 471:\penalty0 87--90, 2011.
\newblock \doi{10.1038/nature09770}.

\bibitem[Wiebe and Granade(2016)]{WIE2016Ga}
Nathan Wiebe and Chris Granade.
\newblock Efficient {B}ayesian phase estimation.
\newblock \emph{Physical Review Letters}, 117:\penalty0 010503, 2016.
\newblock \doi{10.1103/PhysRevLett.117.010503}.

\end{thebibliography}

\end{document}